\documentclass{emulateapj}
\usepackage{epsfig}
\usepackage{amsmath}
\usepackage{natbib}
\usepackage{apjfonts}
\usepackage{graphics}
\newcommand{\beq}{\begin{equation}}
\newcommand{\eeq}{\end{equation}}
\newcommand{\bea}{\begin{eqnarray}}
\newcommand{\eea}{\end{eqnarray}}
\newcommand{\trm}[1]{\textrm{#1}}
\renewcommand{\vec}[1]{\mathbf{#1}}
\newcommand{\sgra}{Sgr A$^\ast$~}

\begin{document}
\title{Stellar Dynamics at the Galactic Center with an Extremely Large Telescope}
\author{
  Nevin N.~Weinberg\altaffilmark{1}, 
  Milo\v s Milosavljevi\'c\altaffilmark{1}, and 
  Andrea M.~Ghez\altaffilmark{2}} 
\altaffiltext{1}{Theoretical Astrophysics, California Institute of Technology, 
Pasadena, CA 91125; nnw, milos@tapir.caltech.edu}
\altaffiltext{2}{Department of Physics and Astronomy, 
University of California, Los Angeles, CA 90095-1562; ghez@astro.ucla.edu}

\begin{abstract}

We discuss physical experiments achievable via the monitoring of
stellar dynamics near the massive black hole at the Galactic center
with a diffraction-limited, next generation, extremely large telescope
(ELT). Given the likely observational capabilities of an ELT and what
is currently known about the stellar environment at the Galactic
Center, we synthesize plausible samples of stellar orbits around the
black hole. We use the Markov Chain Monte Carlo method to evaluate the
constraints that the monitoring of these orbits will place on the
matter content within the dynamical sphere of influence of the black
hole. We express our results as functions of the number $N$ of stars
with detectable orbital motions and the astrometric precision $\delta
\theta$ and spectroscopic precision $\delta v$ at which the stellar
proper motions and radial velocities are monitored. Our results are
easily scaled to different telescope sizes and precisions. For
$N=100$, $\delta \theta = 0.5 \trm{ mas}$, and $\delta v = 10 \trm{ km
s}^{-1}$---a conservative estimate of the capabilities of a 30 meter
telescope---we find that if the extended matter distribution enclosed
by the orbits at 0.01 pc has a mass greater than $\sim 10^3 M_\odot$,
it will produce measurable deviations from Keplerian motion. Thus, if
the concentration of dark matter at the Galactic Center matches
theoretical predictions, its influence on the orbits will be
detectable. We also estimate the constraints that will be placed on
the mass of the black hole and on the distance to the Galactic Center,
and find that both will be measured to better than $\sim 0.1\%$.  We
discuss the significance of knowing the distance to within a few
parsecs and the importance of this parameter for understanding the
structure of the Galaxy. We demonstrate that the lowest-order
relativistic effects, such as the prograde precession, will be
detectable if $\delta \theta \la 0.5 \trm{ mas}$. Barring the
favorable discovery of a star on a highly compact, eccentric orbit,
the higher-order effects, including the frame dragging due to the spin
of the black hole, will require $\delta \theta \la 0.05 \trm{
mas}$. Finally, we calculate the rate at which monitored stars
experience detectable nearby encounters with background stars.  The
encounters probe the mass function of stellar remnants that accumulate
near the black hole.  We find that $\sim 30$ such encounters will be
detected over a ten year baseline for $\delta \theta = 0.5 \trm{
mas}$.

\end{abstract}


\section{Introduction}
\label{sec:intro}

Observational programs with ten meter class telescopes, including the
W.~M.~Keck Observatory and the Very Large Telescope (VLT), have
yielded a wealth of information on the stellar content inside the
sphere of influence of the massive black hole at the Galactic center
(GC;
\citealt{Ghez:98,Gezari:02,Hornstein:02,Figer:03,Genzel:03a,Ghez:03b,Schoedel:03}).
The black hole is located at the center of a compact stellar cluster
that has been the target of observational surveys for a decade
(e.g.,~\citealt{Krabbe:95,Figer:00,Gezari:02}).  Near-infrared
monitoring with speckle and adaptive optics techniques has recently
enabled complete orbital reconstruction of several stellar sources
orbiting the black hole
\citep{Eckart:02,Schoedel:02,Schoedel:03,Ghez:03b}.  Sources have been
monitored with astrometric errors of a few milli-arcseconds
\citep{Ghez:03a,Schoedel:03}, and radial velocity errors $< 50 \trm{
km s}^{-1}$ \citep{Eisenhauer:03, Ghez:03a}, allowing the detection of
the accelerated proper motions of $\sim 10$ stars. One of these stars has an
orbital period of only $\sim 15 \trm{ yr}$ \citep{Ghez:03b, Schoedel:03}.

The presence of a dark mass at the center of the Galaxy could in
principle be inferred from the static nature of the radio source \sgra
located at the center of the stellar cluster
\citep{Backer:99,Reid:99}. Nevertheless, it is the stars with the
shortest orbital periods that have provided unequivocal proof of the
existence of a massive black hole and a measurement of its mass of
$\sim4\times10^6 M_\odot$ \citep{Ghez:03b, Schoedel:03}.  Since,
for a fixed angular scale, the orbital periods are proportional to
$R_0^{3/2} M_{\rm bh}^{-1/2}$ and the radial velocities are
proportional to $R_0^{-1/2}M_{\rm bh}^{1/2}$ where $R_0$ is the
heliocentric distance to the black hole and $M_{\rm bh}$ is its mass,
the two parameters are not degenerate and can be determined
independently \citep{Eisenhauer:03}.

In spite of the quality of elementary data available about the black
hole and the bright stellar sources, the matter content in the
vicinity of the black hole remains unknown.  The observed stellar
sources may represent only a fraction of the total matter content.
Since the radial diffusion time $\sim 10^{8-9}\textrm{ yr}$ is shorter
than the age of the bulge, a large number of massive compact remnants
($5-10M_\odot$ black holes) could have segregated into, and may
dominate the matter density inside the dynamical sphere of influence
of the black hole \citep{Morris:93,Miralda:00}.  Furthermore,
adiabatic growth of the massive black hole could have compressed a
pre-existing distribution of cold dark matter (CDM)
\citep{Ipser:87,Quinlan:95,Gondolo:99} and stars
\citep{Peebles:72,Young:80} into a dense ``spike'' .  A variety of
dynamical processes, however, are capable of destroying such a spike
\citep{Ullio:01,Merritt:02,Gnedin:02,Merritt:03}. A sustained CDM
spike would have implications for the detection of annihilation
radiation for the CDM models in which the CDM consists of
weakly-interacting massive particles (WIMPs).

The most complete catalogue of stars in the central parsecs was
compiled by \citet{Genzel:00} and \citet{Schoedel:03}.  In a survey of
the stellar sources, \citet{Genzel:03a} infer a spatial number density
of $n(r)\propto r^{-1.4}$ over the radial range $0.004<r<0.4\textrm{
pc}$.  Their sample was $50\%$ complete for stars brighter than $K\sim
18$, where completeness is defined as the percentage of stars in the
field of view that are detectable and thus included in the sample.
Expressed in terms of stellar mass, the sample is $50\%$ complete for
masses $m\gtrsim 3M_\odot$, assuming stars on the main sequence, a
distance to the GC of 8.0 kpc \citep{Reid:93} and $K$-band extinction
of 3.3 mag \citep{Rieke:89}. A picture is emerging in which the
brightest stars in the Central Cluster ($<0.03$ pc) are young,
main-sequence stars with apparent magnitudes $K>13$ and masses
$10-15M_\odot$.  The stars outside $0.03$ pc appear to be
spectroscopically and kinematically distinct.  They span a larger
range of magnitudes $K\gtrsim 10$ and contain $\sim40$ mass-losing
Wolf-Rayet stars (e.g.,~\citealt{Genzel:03a} and R.~Genzel, private
communication).  Unlike the Central Cluster, these stars appear to
belong to twin, misaligned stellar disks \citep{Levin:03,Genzel:03a}.

The formation of the observed young stars with $\times100$ larger
specific binding energies relative to the black hole than that of the
nearest observed accumulation of molecular gas
(e.g.,~\citealt{Jackson:93}) presents a challenge to star formation
theories and is a persistent puzzle
(e.g.,~\citealt{Morris:93,Ghez:03a, Genzel:03a}).  A number of mechanisms for the
formation and migration of stars in the tidal field of the massive
black hole have been proposed
\citep{Gerhard:01,Gould:03,Hansen:03,Levin:03,Kim:03,Milosavljevic:04}.
While the mechanisms have important implications, they are also each
deficient in at least one way.

There is a dearth of giants in the GC region \citep{Eckart:95}.
Recently, \citet{Figer:03} measured the radial velocities of 85 cool,
normal giant stars with projected distances from the central region
between $0.1-1$ pc. They find nearly complete deficiency of giants
with large radial velocities ($V_{\rm rad}>200\textrm{ km s}^{-1}$).
Since a star in a circular orbit at a distance of $0.1$ pc from the
black hole has velocity $\sim400\textrm{ km s}^{-1}$, the absence of
any such stars with comparable radial velocities indicates that the
observed giants are indeed limited to the region outside the central
$\sim 0.5$ pc.

While the measured stellar density profile of the Galactic bulge is
consistent with that of a singular isothermal sphere
\citep{Becklin:68}, the profile in the central parsec is not well
known, especially for the lower-mass stellar populations.  Assuming
relaxation that is driven by two-body processes, \citet{Bahcall:76}
showed that the equilibrium phase space distribution for a population
of equal mass stars is a power law in density $\rho\propto
r^{-7/4}$. For a multimass distribution the lighter stars are less
centrally concentrated, resulting in a power-law profile that ranges
from $r^{-3/2}$ for the least massive species to $r^{-7/4}$ for the
most \citep{Bahcall:77,Murphy:91}.  A coeval family of stars in the
central region has reached equilibrium only if it is older than the
relaxation time
\bea
t_E & \sim & \frac{\sigma^3}{G^2 m_{\star} \rho \ln \Lambda} \nonumber \\
    & \approx & 2 \times 10^8 \textrm{ yr} \left(\frac{r}{1\textrm{ pc}}\right)^{1/4} 
                  \left(\frac{M_{\rm bh}}{4 \times 10^6 M_\odot}\right)^{3/2} \nonumber \\
    &  &          \times \left(\frac{m_\star}{10 M_\odot}\right)^{-1} 
                  \left(\frac{\rho_{\rm 1 pc}}{2 \times 10^5 M_\odot \textrm{ pc}^{-3}}\right)^{-1} 
                  \left(\frac{\ln \Lambda}{10}\right)^{-1} 
\eea
where $\sigma$ is the local linear stellar velocity dispersion,
$m_\star$ is the mass of a typical field star, $\rho$ is the local
stellar density, and $\ln \Lambda$ is the Coulomb logarithm.

Since the main sequence lifetime of stars more massive than $\sim 2
M_\odot$ is shorter than $t_E$, young massive stars in the GC are not
relaxed; their distribution is primarily a reflection of their
formative conditions. While lower mass dwarf stars are sufficiently
old to be relaxed in the central potential, their distribution in the
innermost region could be affected by an abundance of stellar mass
black holes ($5 - 10 M_\odot$). As products of normal stellar
evolution, stellar mass black holes sink in the potential of the
massive black hole \citep{Morris:93,Miralda:00} and displace the less
massive stars and remnants.

Speckle imaging and more recently adaptive optics with the Keck and
VLT have provided several milliarcsecond astrometry, enabling the
detection of proper motions within the inner 0.5 pc and accelerated
proper motions of $\sim 10$ stars within the inner 0.05 pc. Radial
velocities with spectroscopic precisions of $\delta v \sim 30 \trm{ km
s}^{-1}$ have also been obtained for the star S0-2, which has been
monitored for over 70\% of its orbit including pericenter passage at
$\sim130\textrm{ AU}$ from the black hole. These observations have
enabled the black hole mass and GC distance to be measured to within
$\sim 10\%$ \citep{Ghez:03b, Schoedel:03}.

Here we examine the extent to which one can probe the GC potential by
monitoring stars with a diffraction-limited, next generation,
extremely large telescope (ELT). As compared with current 10 m class
telescopes, the finer angular resolution of an ELT enables the orbital
motions of many more stars to be detected, each at greater astrometric
precision, $\delta \theta$, and spectroscopic precision, $\delta
v$. Given the range of possible sizes of future telescope and given
the uncertainties in the ultimate capabilities of a specific telescope
class (e.g., 30 meter telescopes) we choose to express our results not
as functions of the ELT aperture but rather as functions of $\delta
\theta$, $\delta v$, and the number $N$ of stars with detectable
orbital motions. We take $\delta \theta = 0.5 \trm{ mas}$ and $\delta v =
10 \trm{ km s}^{-1}$ as our fiducial model, corresponding to a
conservative estimate of the capabilities of a telescope with a $D =
30 \trm{ m}$ aperture. We show that $N$ scales with telescope aperture
as $N \simeq 100 (D /30 \trm{ m})^2$. We demonstrate that with an ELT
one can measure the density profile of a dark matter spike and those
general relativistic effects that scale as $(v/c)^2$, where $v$ is the
speed of a star and $c$ is the speed of light. Furthermore, we show
that the distance to the GC will be measured to remarkable
precision. This will help place tight constraints on models of the
overall Galactic structure. We also show that with an ELT one can
detect the gravitational interactions between monitored stars and the
background massive stellar remnants that accumulate near the central
black hole. Such interactions may probe the mass function of the
stellar mass black holes thought to dominate the matter density in the
region.

The paper is organized as follows. In \S~\ref{sec:observations} we
calculate the number of stars with accelerated proper motions that can
be monitored with a given ELT based on its astrometric,
spectroscopic, and confusion limits. We also describe a realistic
monitoring program and demonstrate that confusion with the infrared
emission from \sgra is unlikely to affect an ELT's ability to measure
stellar motions. In \S~\ref{sec:orbitmodel} we model the orbital data
and estimate the magnitude of various non-Keplerian effects including
Newtonian retrograde precession due to extended matter, relativistic
prograde precession, precession induced by the coupling of orbits to
the spin of the black hole, and the Roemer time delay.  In
\S~\ref{sec:scattering} we consider the effect of stellar interactions
on the motion of the monitored stars. Specifically, we estimate the
rate at which discrete stellar encounters result in detectable changes
of orbital motions. In \S~\ref{sec:method} we discuss a method for
generating mock ELT orbital data and describe a computational
technique for estimating uncertainties in the orbital parameters. The
results of our calculations are given in
\S~\ref{sec:results}. Finally, in \S~\ref{sec:discussion} we discuss
astrophysical applications of the proposed observations.

\section{Observing Stars in the Central Arcsecond with an ELT}
\label{sec:observations}
The purpose of this section is to estimate the number and distribution
of stars whose orbital motions can be detected with an ELT and to
determine the astrometric and spectroscopic precision to which their
motions can be measured (\S~\ref{sec:astrometry}).  The latter are
determined by the specifications of the telescope and the properties
of the stellar population at the GC.  Several factors complicate the
monitoring of orbits within the central arcsecond. The greatest
obstacle to detecting and following hitherto unseen stars is stellar
crowding.  Light contamination from nearby bright stars as well as the
light from underlying faint stars flood the pixel elements and impose
a limit to the faintest detectable star.  In \S~\ref{sec:confusion} we
estimate the minimum luminosity permitted by the crowding and thereby
obtain an estimate of the number count of stars with observable
orbital motions.

\subsection{Astrometric and Spectroscopic Limit}
\label{sec:astrometry}
With adaptive optics, an ELT will operate near its diffraction limit
in the $K$-band. By determining the centroid of images, the measured
relative positions of stars are a factor of $\sim 20-40$ more definite
than the images' diffraction limit. For instance, the diffraction
limit of Keck is $\sim 50 \textrm{ mas}$ while the astrometric error
of a bright star near the GC as seen by Keck is $\sim1-2 \textrm{
mas}$. Naively, the expected astrometric limit of an ELT with $D = 30
\trm{ m}$ is therefore $\delta \theta_{30} \sim 0.5 \textrm{ mas}$.

In practice, the astrometric limit achievable with adaptive optics
depends on whether atmospheric fluctuations or centroid measurement
errors dominate the signal. At the GC the separation between the guide
star needed for the adaptive optics infrared wavefront sensor and the
star under study is typically $\sim 5\arcsec$, corresponding to a
separation of $0.25\textrm{ m}$ at the top ($\sim 10 \textrm{ km}$) of
the atmosphere.  As long as the telescope aperture is larger than this
separation, as is the case for Keck and an ELT, the atmosphere
dominates and the astrometric precision scales with the telescope
diameter as $D^{2/3}$ \citep{Shao:92}. A 30 meter ELT is expected to
have an astrometric limit that is $3^{2/3}\approx2$ times smaller than
Keck's for $K\lesssim 24$. We therefore adopt $\delta \theta_{30} \sim
0.5 \textrm{ mas}$ in our calculations, though we consider this a
conservative estimate; a 30 meter ELT may attain an astrometric limit
as small as 0.1 mas.\footnote{For example, see
http://tmt/ucolick.org/reports\_and\_notes/index/htm, Report No. 34.}

With an adaptive-optics-fed spectrometer on Keck, \citet{Ghez:03a}
detected spectral absorption lines in the star S0-2 at a spectral
resolution of $R = \lambda / \Delta \lambda \sim 4000$, yielding a
radial velocity measurement with an error of $40 \trm{ km s}^{-1}$
(see also \citealt{Eisenhauer:03}). Integral field spectroscopy in the
near-IR with a 30 meter ELT is expected to enable measurements with $R
\sim 1 - 2 \times 10^4$, suggesting that velocity errors of $\delta
v_{\rm 30} \sim 10 \trm{ km s}^{-1}$ are attainable. This too is a
conservative estimate as a 30 meter ELT may achieve velocity errors
more than an order of magnitude smaller (D. Figer, private
communication). As we discuss below, an ELT will be able to detect
stars that are fainter and hence cooler than those currently
detectable. Cool stars exhibit rich spectral features including
possible molecular lines, enabling high spectral resolution
studies. For example, \citet{Figer:03} obtained radial velocities for
85 cool stars in the central parsec of the Galaxy with velocity errors
of $\sim 1 \trm{ km s}^{-1}$.

Although an ELT's astrometric and spectroscopic limits may differ from
the above estimates, we show in \S~\ref{sec:mockdata} that the
uncertainties in the model parameters extracted from the monitoring
data, such as the distance to the GC and the extended matter profile,
scale almost linearly with the measurement errors. The constraints on
the parameters for different values of $\delta \theta$ and $\delta v$
can therefore be readily inferred from our results.

\subsection{Confusion Limit}
\label{sec:confusion}

The brighter stars wash out the signal of fainter stars, thereby
limiting the luminosity of the faintest observable star. This limit
depends on the telescope optics (e.g., angular resolution) and on the
stellar luminosity function (LF).  Using measurements of stellar
photometry near the GC, we now estimate the minimum luminosity that a
star at the GC can have and still be identified and monitored with an
ELT.  For a given star of luminosity $l$ and for a given $K$-band
stellar LF, we determine the integrated flux from all nearby
background stars with luminosity $<l$. At some minimum luminosity, the
emission from a single star is comparable to the background emission;
this luminosity sets the confusion limit.

\begin{figure*}
\begin{center}
\includegraphics[angle=-90, scale=0.65]{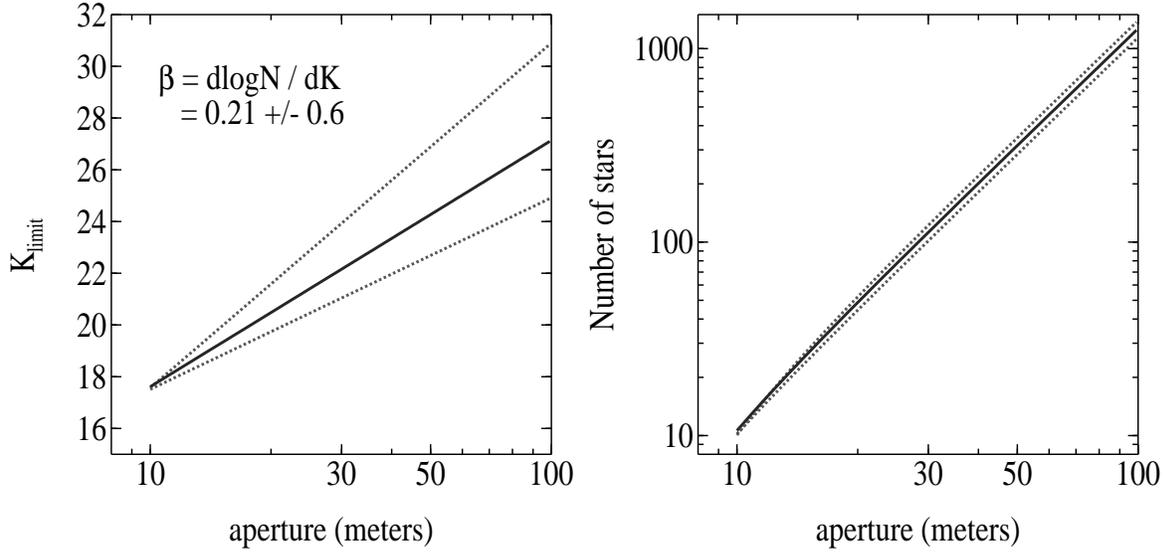}.
\caption{ The $K$-band magnitude limit and number of stars with
detectable orbital motions as a function of the aperture of a
diffraction limited ELT. Results are shown for power-law $K$-band
luminosity functions normalized to observations by \citet{Schoedel:03}
with slopes matching the $\sim 2 \sigma$ range found by
\citet{Genzel:03a}. The uncertainty in the number counts is
considerably smaller than the uncertainty in the magnitude
limits. \label{fig:Klimit}}
\end{center}
\end{figure*}

Following \citet{Takeuchi:01} and references therein, let $x_S = S\,
h(\theta, \phi)$ be the response of the telescope to a source of flux
density $S$ at an angular position $(\theta, \phi)$ from the
line-of-sight axis to the center of the source. $h(\theta, \phi)$ is
the point-spread function (PSF) of the telescope, normalized to unity
at the center. Since all sources at the GC are essentially at the same
distance, we can instead express the response in terms of stellar
luminosity $l$, i.e., let $x = l\, h(\theta, \phi)$.  The variance in
the telescope response due to crowding is the confusion noise
$\sigma$. To detect a source with high statistical significance, its
luminosity must be greater than some cutoff $l_c$, or equivalently,
$x$ must be greater than a response cutoff $x_c$. Defining $q = x_c /
\sigma$, a source is above the confusion limit if its signal-to-noise
$S = x / \sigma > q$, where we take $q =5$.

If the number of stars per square arcsec with luminosity in the range
$(l, l + dl)$ is $dN = \alpha \Phi(l)dl$, where $\alpha$ is the
normalization of the LF and $\Phi(l)$ is its shape, then the mean
number of source responses of intensity $x$ is
\bea
R(x) dx & = & \int \frac{dN(l)}{dl} dl d\Omega \nonumber \\
        & = & \int \alpha \Phi\left(\frac{x}{h(\theta, \phi)}\right) 
              \frac{d\Omega}{h(\theta, \phi)} dx,
\eea
where the integral is over the solid angle of the PSF.  The confusion
noise $\sigma$ due to all sources fainter than $x_c$ is then
\beq
\sigma^2 = \int_0^{x_c} x^2 R(x) dx .
\eeq 

Since we are interested in calculating the cutoff response of a given
detector for a given LF, we need to solve for the confusion noise.
Assuming a power law LF of the form $dN/dl = \alpha \Phi(l) = \alpha
l^{-\eta}$ we have
\bea
R(x) & = & \int \alpha \left[\frac{x}{h(\theta,\phi)}\right]^{-\eta} 
                                                      \frac{d\Omega}{h(\theta, \phi)} 
= \alpha x^{-\eta} \Omega_{\rm eff} ,
\eea
where
\beq
\Omega_{\rm eff}  =  \int h(\theta,\phi)^{\eta - 1} d\Omega.
\eeq
Therefore
\bea
\sigma^2 & = & \alpha \Omega_{\rm eff} \int_0^{x_{\rm c}} x^{2 - \eta} dx
=\left(\frac{q^{3-\eta}\alpha \Omega_{\rm eff}}{3 - \eta} \right)^{2/(\eta-1)} . 
\eea
In this paper we only consider power-law LFs, though one can obtain an
expression for $\sigma$ for general forms of the LF
\citep{Franceschini:89}.

For a Gaussian PSF $h(\theta, \phi) = h(\theta) = \exp [-(4 \ln
2)(\theta / \theta_0)^2 ]$, where $\theta_0$ is the PSF's full-width
at half-maximum.  This gives $\Omega_{\rm eff} = \pi
\theta_0^2/\left[(4 \ln 2)(\eta - 1)\right]$ and thus
\beq
\label{eq:sigma}
\sigma =  \left(\frac{q^{3-\eta}}{3 - \eta} \right)^{1/(\eta-1)}
         \left[\frac{\pi \theta_0^2 \alpha}{(4 \ln 2)(\eta - 1)}\right]^{1/(\eta-1)} .
\eeq
 
We now estimate the value of $\sigma$ for the Keck and an ELT. In the
$K$-band, $\theta_0 \simeq 50 \trm{ mas}$ for Keck and $\theta_0
\simeq 15 \trm{ mas}$ for a $D = 30 \trm{ m}$ ELT.  We also need the
$K$-band luminosity function (KLF) of stars at the
GC. \citet{Genzel:03a} find that the KLF within $1.5 \arcsec$ of the
GC is well described by a power-law with slope $\beta = d \log N / dK
= 0.21 \pm 0.03$ where $K$ is the apparent magnitude in the
$K$-band. We consider KLFs with slopes within the $\sim 2\sigma$ range
$0.15 < \beta < 0.27$, which in terms of $\eta = 1 + \beta/0.4$
corresponds to the range $1.38 < \eta < 1.68$.

 \citet{Schoedel:03} measured the photometry of more than 40 stars in
the central arcsec, 29 of which reside within $0.8\arcsec$ ($\sim6000
\trm{ AU}$). We normalize the KLF to these 29 stars. We limit our
analysis to these innermost stars since the KLF inside $0.8\arcsec$
appears to differ from that outside this region (see
\S~\ref{sec:intro}). We do not attempt to account for a possible
radial dependence but instead assume the KLF is constant.

Of the 29 stars, the brightest has apparent magnitude $K = 13.4$ and
the dimmest $K = 17.3$. Assuming a $K$-band extinction of 3.3 mag
\citep{Rieke:89} and a distance to the GC $R_0 = 8 \trm{ kpc}$, these
apparent magnitudes correspond to $K$-band luminosities of $l_{\rm
min} = 0.8 L_\odot$ and $l_{\rm max} = 28 L_\odot$. For a given $\eta$
we calculate $\alpha = N_{\rm obs} (1 - \eta) / (l_{\rm max}^{1-\eta}
-l_{\rm min}^{1-\eta})$ where $N_{\rm obs}=29 / \pi (0.8\arcsec)^2 $ and
by equation (\ref{eq:sigma}) solve for $\sigma$. Integrating the
luminosity function over stars brighter than $x_c = q \sigma$ yields
the number count of detectable stars $N(l > l_c) \sim \alpha
x_c^{1-\eta} / (\eta - 1)$.

In Figure \ref{fig:Klimit} we show how the $K$-band magnitude limit
and number $N$ of stars with detectable orbital motions (those within
3000 AU of the GC; see below) scale with the aperture of a diffraction
limited ELT assuming $\beta = 0.21 \pm 0.06$ ($\eta = 1.53 \pm
0.15$). Since by equation (\ref{eq:sigma}) $N \propto x_c^{1-\eta}
\propto \theta_0^{-2}$, we find that $N \simeq 100 (D / 30 \trm{
m})^2$. Furthermore, because $x_c \propto \alpha^{1 / (\eta - 1)}$,
$N(l > l_c)$ is not very sensitive to the value of $x_c$ for a fixed
$\eta$. Therefore, when the above analysis is performed on a subset of
the 29 stars within $0.8 \arcsec$ (e.g., stars within $0.4 \arcsec$ or
alternatively stars brighter than $K = 16$), the derived number
counts, unlike the magnitude limits, do not change significantly. The
number counts we derive for an ELT are therefore robust even though
the magnitude limits are subject to some uncertainty.

To extract orbital parameters the acceleration of a star in the plane
of the sky must be detected, i.e., it must be greater than the
threshold acceleration $\xi_t$.  For Keck $\xi_t \sim 1 - 2 \trm{ mas
yr}^{-2}$ while for a 30 meter ELT $\xi_t \sim 0.5 \trm{ mas
yr}^{-2}$. The accelerated proper motion is detectable over the entire
orbit if the acceleration at apocenter exceeds the threshold. For a
face on orbit this requires $a(1+e) < (GM/\xi_t R_0)^{1/2}$. Thus the
acceleration will be detectable with a 30 meter ELT over the entire
orbit if $a \la 3000 \trm{ AU}$ (period $\la 80 \trm{ yr}$). To
construct our mock stellar orbits to simulate observations that can be
made with an ELT, we only consider orbits satisfying this
constraint. As Figure \ref{fig:Klimit} shows, within 3000 AU,
approximately 100 stars are brighter than a 30 meter ELT's confusion
limit. Furthermore, since the surface density of stars is $\sim 200
\trm{ arcsec}^{-2}$, one does not expect to observe stars with
apocenters smaller than $\sim 300 \trm{ AU}$ ($\sim 0.04\arcsec$) with
such an ELT.  We therefore conclude that a 30 meter ELT will detect
the accelerated motion of $\sim100$ stars with semi-major axes $\la
3000 \trm{ AU}$ and apocenter distances $\ga 300 \trm{ AU}$. 

Another related issue is the frequency with which an ELT will measure
positions and radial velocities for the $N$ monitored stars, given a
reasonable commissioning of $\sim10$ GC exposures per year. As it will
be equipped with an integral field unit spectrometer an ELT can
obtain simultaneous spectral and spatial data over a relatively large
region of sky. It is possible for it to measure positions and
velocities for all $N$ stars in a single image. This suggests that a
dedicated observing program can reasonably obtain ten measurements per
star per year.

At the high levels of precision obtainable with an ELT, orbital
parameter constraints should scale with the measurement errors
$\sigma$ (i.e., $\delta \theta$ and $\delta v$) and the number of
stars $N$ as $\sigma / N^{1/2}$. Based on the above discussion we
therefore expect the parameter constraints to scale with telescope
aperture as $\sim D^{-2/3} / D \sim D^{-5/3}$. We verify this relation
in our numerical simulation results described in \S~\ref{sec:results}.

\subsection{Central Point Source --- \sgra}

At radio wavelengths \sgra is detected as a nonthermal
\citep{Beckert:96, Serabyn:97}, compact \citep{Rogers:94}, static
\citep{Backer:99, Reid:99}, variable \citep{Zhao:01} source.  An X-ray
source coincident with \sgra has also been detected
\citep{Baganoff:03} and consists of a resolved, steady-state
component with size $\sim1\arcsec$ and an unresolved flaring component
that increases in flux density by an order of magnitude over the
course of a few hours roughly once per day
\citep{Baganoff:01,Goldwurm:03,Porquet:03}.

Recently, a near-infrared counterpart to \sgra, located within a few
mas of the dynamically determined black hole position, has been
observed in the $H$ band ($1.7 \mu\trm{m}$; \citealt{Genzel:03b}) and
the $L^\prime$ band ($3.8 \mu\trm{m}$; \citealt{Ghez:04}). Like the
X-ray emission, the infrared emission consists of a quiescent
component and a variable component. The latter exhibits flux densities
that increase by a factor of a few over the course of tens of minutes
to one week, with possible signs of periodicity
\citep{Genzel:03b}. The observed $L^\prime$-magnitudes are in the
range $12.2-13.8$, corresponding to $L^\prime$-luminosities $\sim
10-100L_\odot$.

Although the current sample of stars with detected accelerated motion
are brighter than the \sgra infrared emission, the stars detectable
with an ELT will have comparable luminosities.  A star that passes
near the black hole can therefore be confused with the emission of
\sgra. Conservatively, such confusion limits monitoring when the
projected separation between a star and \sgra is smaller than the
resolution of the detector. For a 30 meter ELT operating at the
diffraction limit in the $K$ band, this corresponds to $\sim 15 \trm{
mas}$ (120 AU). However, as we found in \S~\ref{sec:confusion},
confusion with nearby stars precludes such a telescope from detecting
orbits with apocenters smaller than 300 AU ($\sim 40 \trm{ mas}$) and
most of the monitored stars do not therefore pass within 15 mas of
\sgra. Of those that do, most spend only a small fraction of their
total orbital period that close to the black hole; e.g., a star with a
semi-major axis of 200 AU and eccentricity 0.9 is within 15 mas of the
black hole for only 10\% of its orbital period.  The same arguments
hold for other ELT apertures. Therefore, the infrared emission from
the black hole will not significantly impair orbital monitoring with
an ELT.

\section{Orbital Dynamics}
\label{sec:orbitmodel}
While current observations of stellar proper motions near the black
hole at the GC are consistent with motion around a Newtonian point
mass, we show that with an ELT non-Keplerian motions are going to be
detectable. There are various effects that cause deviations from
Keplerian motion, including the Newtonian retrograde precession (NRP)
of an orbit due to the presence of an extended matter distribution
(\S~\ref{sec:newtonian}), the relativistic prograde precession (RPP;
\S~\ref{sec:relativistic}), and the frame dragging effects related to
the black hole spin (\S~\ref{sec:spin}).  In addition, we account for
an apparent deformation of the observed proper motion (``Roemer
effect''; \S~\ref{sec:roemer}) due to the differences in light travel
times at different locations along the orbit.  A discussion of the
effects of encounters between monitored stars and background stars is
given in \S~\ref{sec:scattering}. We now describe the orbital
equations of motion and estimate the magnitude of the various
non-Keplerian effects. A number of relativistic effects, including
those we consider below, are discussed in \citet{Pfahl:03} in
connection with long-term timing observations of a radio pulsar that
might be detected in a $\la 100$ year orbit about the GC.

\subsection{Equations of Motion}
\label{sec:EOM}
We found in \S~\ref{sec:observations} that we do not expect a 30 meter
ELT to detect orbits with apocenter smaller than $\sim 300 \trm{ AU}$
due to confusion noise. Assuming orbits uniformly distributed in
$e^2$, the probability that a given star has $e > 0.99$ is
$2\%$. Since most of the $\sim 100$ stars such an ELT monitors will
have semi-major axes $>1000\textrm{ AU}$, it is unlikely that any will
have pericenter distance smaller than a few AU.  As a result, the
ratio of the Schwarzschild radius to the pericenter distance of the
stars will satisfy $R_s / r_p \la 0.05$, or expressed in terms of the
stellar velocity at pericenter, $v_p / c \la 0.2$.  The post-Newtonian
approximation to the geodesic equations that is accurate to order
$(v/c)^2$ provides an adequate description of the stellar orbits given
the observational precision expected with an ELT.
 
The geodesic equation for test particles orbiting a spherically
symmetric mass is, in the post-Newtonian approximation
\citep{Weinberg:72,Rubilar:01},
\bea
\label{eq:EOM}
\frac{d\vec{v}}{dt} &=& -\nabla\Phi \nonumber \\ & & - \left[2\nabla\Phi^2
+ v^2 \nabla \Phi - 4\vec{v}(\vec{v} \cdot \nabla)\Phi -
\vec{v} \times (\nabla \times \vec{\zeta}) \right]/c^2, 
\eea
where $\vec{v} = d \vec{x}/d t$ is the velocity vector, $\Phi$ is a
time-independent gravitational potential, and
$\vec{\zeta}=2G({\vec{x}}\times{\vec{J}})/r^3$ is a vector potential
associated with the spin $\vec{J}$ of the gravitating mass, which we
assume is constant with time.  We assume the density distribution of
the extended matter at radii traversed by the stars and smaller is a
power-law profile $\rho(r)=\rho_0(r/r_0)^{-\gamma}$.  Input model
parameters are described in \S~\ref{sec:mockdata}.

The relativistic effects include corrections to the orbital dynamics
and to the observed motion due to propagation effects. The former
class includes the RPP and frame dragging while the later class
includes the lowest order ($v/c$) Roemer time delay and such $(v/c)^2$
effects as time dilation, gravitational redshift, and the Shapiro time
delay. Since the $(v/c)^2$ propagation effects each have different
functional dependences on the orbital parameters (see e.g.,
\citealt{Pfahl:03}), including them may break degeneracies, though
they may also weaken the sensitivity to some parameters. However, our
interest in the $(v/c)^2$ relativistic effects is primarily connected
with the ability of an ELT to probe general relativity on the scale
of a massive black hole rather than with the effects' potential
utility for parameter estimation. Since $v_p/c < 0.2$, relativistic
effects help to constrain the orbital parameters by at most a few
percent. We therefore chose not to include the $(v/c)^2$ propagation
effects in our analyses. The analysis of actual data obtained from an
ELT must, however, account for all the relativistic effects.

\subsection{Newtonian Retrograde Precession}
\label{sec:newtonian}

The NRP was discussed in the context of the GC by \citet{Rubilar:01}.
An extended matter distribution causes stellar orbits to precess due
to differences in the amount of mass that is contained between the
apocenter and the pericenter radii.  In the numerical calculations
that follow, we determine how much an orbit precesses due to the
extended matter by solving equation (\ref{eq:EOM}).  Here, however, we
obtain an estimate of the magnitude of the precession for stars at the
GC by considering the potential of the extended matter to be a small
correction $\delta \Phi$ to the potential of the black hole.  Expanding the 
total potential to linear order in $\delta \Phi$, the angular shift per 
period is \citep{Landau:60}
\beq
\Delta \phi_{\rm Newt} = \frac{\partial}{\partial L}
                         \left(\frac{2}{L} \int_0^{\pi} r^2 \delta \Phi \; d\varphi \right),
\eeq
where $L$ is the orbital angular momentum per unit mass, $r = a(1-e^2)
/ (1 + e \cos \varphi)$, and $\varphi$ is the phase of the orbit.  If
$\gamma<2$ we have $\delta \Phi=\beta r^{2-\gamma}$, where $\beta = G
M_{\rm ext} /(2-\gamma) r_0^{3-\gamma}$ is a constant and 
$M_{\rm ext} \equiv M_{\rm ext}(r<r_0)$ is the extended matter mass 
within $r_0$.  The orbital precession is then given by \citep{Munyaneza:99}
\beq
\label{eq:Newtangle}
\Delta \phi_{\rm Newt} = \frac{2 \beta}{G M_{\rm bh}} 
\left[a(1-e^2)\right]^{3-\gamma} g(\gamma,e),
\eeq
where
\bea
g(\gamma, e) &=& \frac{1-e^2}{e^2}(4-\gamma)
\left[I_{4-\gamma}(e)-I_{5-\gamma}(e)\right] \nonumber \\ & & + 
   (7-2\gamma) I_{4-\gamma}(e) 
\eea
and
\beq
I_{n}(e) \equiv \int_0^\pi \frac{d\varphi}{(1+e \cos \varphi)^n} .
\eeq

Assume that the extended matter consists of stars with $\gamma=7/4$
and $\rho_{\rm 1 pc} = 2 \times 10^5 M_\odot\trm{ pc}^{-3}$ (see
\S~\ref{sec:intro}).  Consider an S0-2 like orbit with a semi-major
axis of $0.005 \trm{ pc}$ and eccentricity $e = 0.9$.  The enclosed
stellar mass at apocenter and pericenter are $6000 M_\odot$ and $150
M_\odot$. Solving equation (\ref{eq:Newtangle}) yields a precession
per revolution of $\Delta \phi_{\rm Newt} \approx 0.08^\circ$,
corresponding to an apparent angular apocenter shift of roughly $
\Delta \phi_{\rm Newt} a (1+e)/ R_0 \approx 0.3 \trm{ mas}$. Thus, a
few S0-2 like orbits with astrometric errors of 0.5 mas provide a
meaningful constraint on the stellar distribution within the inner few
milliparsecs. If the density of the dark matter cusp at the stellar
positions exceeds $\sim 10^8 M_\odot \trm{ pc}^{-3}$, then it too will
produce a detectable precession; it will not be easily distinguished
from the stellar contribution (however see \S~\ref{sec:scattering}).

\subsection{Relativistic Prograde Precession}
\label{sec:relativistic}

The RPP causes a pericenter advance per revolution of (see \citealt{Weinberg:72}) 
$\Delta \phi_{\rm pro} = 3 \pi R_s/a (1 - e^2)$, where $R_s=2GM_{\rm bh}/c^2$
is the Schwarzschild radius of the black hole. The magnitude of the
effect is $\propto(v/c)^2$.  The apparent apocenter shift per
revolution caused by the RPP is $\Delta s \approx \Delta \phi_{\rm pro} a
(1+e) / R_0 = 3 \pi R_s/ R_0 (1 - e)$, which corresponds to an
apparent shift of $\sim 1 \trm{ mas}$ for the star S0-2. Although the
RPP has an additional factor of $(v/c)$ relative to the Roemer effect
(\S~\ref{sec:roemer}), this attenuation can be compensated by having a
few high eccentricity stars in the sample. Furthermore, unlike the
Roemer effect, the RPP shift is to first order independent of the
semi-major axis and is therefore equally sensitive to stars at all
radii (although stars at large radii also have long periods). Consider
an orbit seen face on and observed for $N_{\rm orb}$ complete
periods. Since the precession angles per revolution add linearly, the
signal-to-noise from the RPP is 
$\emph{S}_{\rm pro} \sim\Delta s N_{\rm orb}/\delta \theta$, or
\bea
\emph{S}_{\rm pro} & \sim & 0.1 \frac{N_{\rm orb}}{1-e} 
\left( \frac{M_{\rm bh}}{4 \times 10^6 \; M_\odot}\right)
                              \left(\frac{R_0}{8 \; \rm{kpc}} \right)^{-1}
                              \left(\frac{\delta \theta}{1 \; \rm{mas}} \right)^{-1}. \nonumber \\ & &  
\eea
In a sample of 100 stars observed with astrometric errors of $0.5
\trm{ mas}$ and having an eccentricity distribution uniform over
$e^2$, we expect on average eight stars with $e > 0.96$. If only one
such a star is followed over just a single period, the RPP shift will
be measured to 5-$\sigma$ accuracy.

\subsection{Frame Dragging}
\label{sec:spin}

For a spinning black hole, frame dragging effects also cause a
precession of the pericenter. The spin precession per revolution for a
star orbiting a black hole with spin angular momentum $J$ is given
approximately by (see \citealt{Weinberg:72}, equation (9.5.22); note different notation)
\bea
\Delta \phi_{\rm spin} & \approx & -8 \pi j \left(\frac{G M_{\rm bh}}{c L}\right)^3 \cos \psi \nonumber \\
 & = & - \frac{2 \sqrt2 j  \Delta \phi_{\rm pro}}{3} \sqrt{\frac{R_s}{a (1-e^2)}} \cos \psi,
\eea
where $\psi$ is the angle between the orbital angular momentum vector
and the black hole spin axis and $0 \leq j \equiv cJ / GM_{\rm bh}^2
\leq 1$ is the black hole spin parameter.

The black hole spin induces an apocenter shift that is smaller than
the RPP shift by a factor of $\sim v/c$. Even if the black hole is
maximally spinning ($j=1$), the shift represents only a $5\%$
contribution on top of the RPP for a star with $a = 200 \trm{ AU}$ and
eccentricity $e = 0.92$. For an orbit observed face-on the
signal-to-noise from a spin-induced apocenter shift is
\bea
\label{eq:SNspin}
\emph{S}_{\rm spin}  & \approx &  \frac{2 \pi  \sqrt{2}j}{R_0 \sqrt{a (1+e)}} \left(\frac{R_s}{1-e}\right)^{3/2} 
         \frac{N_{\rm orb}}{\delta \theta} \cos \psi \nonumber \\
                   & \approx & 0.001 \frac{j N_{\rm orb} \cos \psi}{\sqrt{(1+e)(1-e)^3}} \left( \frac{M_{\rm bh}}{4 \times 10^6 \; M_\odot}\right)^{3/2}  
		\nonumber \\ & & \times
                	          \left( \frac{a}{1000 \trm{ AU}}\right)^{-1/2}
                              \left(\frac{R_0}{8 \; \rm{kpc}} \right)^{-1}
                              \left(\frac{\delta \theta}{1 \; \rm{mas}} \right)^{-1}. 
\eea
For example, a 5-$\sigma$ detection is achieved with an ELT with
$\delta \theta = 0.5 \trm{ mas}$ if a star with $a = 300 \trm{ AU}$
and $e=0.99$ is monitored for three complete orbits. We expect a 30
meter ELT to detect one star with a semi-major axis that small
(\S~\ref{sec:observations}). Assuming eccentricities uniformly
distributed in $e^2$ the probability that star has $e > 0.99$ is only
$\sim 2 \%$. If $\delta \theta = 0.05 \trm{ mas}$, a star with $a =
300 \trm{ AU}$ and $e=0.95$ will yield a 5-$\sigma$ detection after
being monitored for three complete orbits. Since so high a resolution
requires an ELT with aperture $D \sim 100 \trm{ m}$, there will be several
stars with $a \la 300 \trm{ AU}$ and of these $\sim 10\%$ will
have $e > 0.95$. Detecting such a stellar orbit is therefore not unlikely.  
The spin-induced orbital precession thus requires an astrometric
precession of $\delta \theta \la 0.05 \trm{ mas}$ (see also
\citealt{Jaroszynski:98,Fragile:00}).

\subsection{The Roemer Time Delay}
\label{sec:roemer}

For orbits with non-zero inclination, the distance between the Earth
and star, and hence the difference in time between stellar emission
and observation, varies with orbital phase.  This time delay, given
by $\Delta t = t_{\rm{obs}} - t_{\rm{em}} = z(t_{\rm{em}}) / c$, where
$z(t)$ is the relative distance between the star and the massive black
hole, was first recognized by Roemer in 1676 in application to the
phases of Jupiter's moons. Unlike the relativistic Doppler effect
which includes corrections of order $(v/c)^2$ and higher, the Roemer
delay is the classical Doppler effect which only includes terms up to
order $v/c$ (see, e.g.,~\citealt{Loeb:03}).  The delay has a magnitude
corresponding to a few percent of a year for an S0-2 like orbit, and
is observed as an additional shift in the apparent stellar position
with time, $\Delta s(t)$. For a circular orbit seen edge-on the
stellar positions $z(t)$ and $s(t)$ are sinusoidal so that
\bea
\Delta s(t) / a & = & \cos(\omega t_{\rm{obs}}) - \cos(\omega t_{\rm{em}}) \nonumber \\
& = & \cos[\omega(t_{\rm{em}} + a/c \sin(\omega t_{\rm{em}}))] - 
\cos(\omega t_{\rm{em}}) \nonumber \\
& \simeq & - \frac{\omega a}{c} \sin^2(\omega t_{\rm{em}}),
\eea
where $\omega = 2 \pi / P$ and we used the fact that for orbits at the
GC $v\ll c$. The maximum shift, in units of the semi-major axis, is
therefore $v / c$. For non-zero eccentricity and arbitrary
inclination the star's projected position and distance as a function
of time are (see e.g., \citealt{Murray:99})
\beq
\left\{ \begin{array}{c} x(t) \\ y(t) \\ z(t) \end{array} \right\} =
r(t) \left\{ \begin{array}{c} 
\cos\left[\varphi(t) + \alpha \right] \\ 
\sin\left[ \varphi(t) + \alpha \right]\cos i \\ 
\sin\left[\varphi(t) + \alpha \right] \sin i \end{array} \right\}
\eeq
where $i$ is the inclination, $\varphi(t)$ is the orbital-phase (i.e.,
the true anomaly), $\alpha$ is the argument of pericenter, and we
chose the reference direction so that the $x$-axis coincides with the
longitude of ascending node. The Roemer shift is then $| \Delta s | =
\left(\Delta x^2 + \Delta y^2 \right)^{1/2}$ where $\Delta x =
x(t_{\rm obs}) - x(t_{\rm em})$ and similarly for $\Delta y$. To
linear order in $v/c$ the orbit-averaged Roemer shift can be written
as
\beq
\langle |\Delta s(t)| / a \rangle \simeq \frac{\omega a}{c} \sin i \sqrt{f(e, i, \alpha)}
\eeq
where $f(e, i, \alpha)$ is a factor of order unity. For the two extreme cases
$\alpha = 0$ and $\alpha = \pi /2$ (corresponding to the line-of-node
along the major-axis and minor-axis, respectively), $f(e, i, \alpha)$ is
given by
\bea
f(\alpha=0) &=&  \frac{\sqrt{1-e^2}}{2e^4} \left\{2e^4 \cos^2 i \left(3 - \sqrt{1-e^2}\right) \right.\nonumber \\ & &  \left.
+ 3 e^2 \left[2 - \sqrt{1-e^2} - \cos^2 i \left(4- 3 \sqrt{1-e^2} \right) \right]
  \right. \nonumber \\ & & \left. - 6 \sin^2 i \left(1- \sqrt{1-e^2}\right) \right\}, \\
f(\alpha=\pi/2) &=&  \frac{\sqrt{1-e^2}}{2e^4} \left\{-2e^4 \left(2 - \sqrt{1-e^2}\right)  \right.\nonumber \\ & &  \left.
  + e^2 \left[10 - 7 \sqrt{1-e^2} - \cos^2 i \left(4- \sqrt{1-e^2} \right) \right] \right.
 \nonumber \\ & & \left. - 6 \sin^2 i \left(1- \sqrt{1-e^2}\right) \right\}.
\eea

An ELT will be able to detect the effect of the Roemer delay in
orbits at the GC. The signal-to-noise from $N_{\rm obs}$ observations
of an orbit measured with astrometric errors $\delta \theta$ is
approximately $\emph{S}_{\rm delay} \sim \langle \Delta s \rangle
N_{\rm obs}^{1/2}/ R_0 \delta \theta$, or
\bea
\emph{S}_{\rm delay} & \approx &  0.8 \sqrt{N_{\rm obs}} \sin i \; 
\sqrt{f(e, i, \alpha)}  \left(\frac{a}{1000 \; \rm{AU}} \right)^{1/2} 
 \nonumber \\
         & &   \times   \left( \frac{M_{\rm bh}}{4 \times 10^6 \; M_\odot}\right)^{1/2}
            \left(\frac{R_0}{8 \; \rm{kpc}} \right)^{-1} \left(\frac{\delta \theta}{1 \; \rm{mas}} \right)^{-1}.
\eea
If, e.g., we pick $i\sim \pi/3$, $\alpha = 0$, and $e\sim 1 /
\sqrt{2}$, an astrometric error of $0.5 \trm{ mas}$, a mean semi-major
axis of $1000 \trm{ AU}$, and 10 observations per star, then we can
detect the delay to $\emph{S}_{\rm delay} \sim 5$ with roughly 10
stars.  We therefore expect the Roemer delay to be detectable in such
an ELT's sample of $\sim 100$ stars and the effect must be taken into
account during parameter estimation.

\subsection{Interstellar Interactions}
\label{sec:scattering}
In the previous sections we described the motion of a star in the
potential of a black hole and a smooth distribution of extended
matter, including stars, remnants, and dark matter.  This
approximation ignores the fact that the potential due to stars and
remnants is the sum of discrete point-mass potentials and is therefore
not perfectly smooth. The stars experience perturbations due to nearby
encounters with individual stars and due to fluctuations in the
potential arising from all stars. These perturbations cause a star's
orbital parameters to change with time.  The magnitude and the rate of
these changes depend on the stellar mass function since the
perturbations are sensitive to the characteristic mass of the field
stars.  Thus, measuring the effects of stellar encounters is a probe
of the mass function in the central parsec.  It also breaks the
degeneracy between the contributions of stellar matter and dark matter
to the Newtonian orbital precession. Encounters may also be a source
of noise in measurements of orbital parameters such as the black hole
mass and distance to the GC. While we do not include the effects of
encounters in our numerical calculations presented in
\S~\ref{sec:results}, we now estimate their magnitude and demonstrate
that the encounters might be detectable with an ELT and present a
powerful probe of the mass function of stellar remnants at the GC.

An encounter between a test star of mass $m_j$ and a field star of
mass $m_i$ with impact parameter $b$ induces a change in the test star
velocity given by (see, e.g., \citealt{Spitzer:87})
\bea
\label{eq:deltav}
\delta v= \frac{2 m_i v_{\rm rel}}{m_i + m_j} \left[ 1 + \left(\frac{b}{b_0}\right)^2 \right]^{-1/2} ,
\eea
where $b_0 = G (m_i + m_j) / v_{\rm rel}^2$ and $v_{\rm rel}$ is the
initial relative velocity of the stars. The encounter induces a change
in the test star's velocity distinct from that due to orbital motion
around the black hole. We solve for the maximum impact parameter
$b_{\rm max}$ such that an encounter induces a change in velocity of
the test star larger than the minimum detectable change $\delta v_{\rm
min}$.  For uncorrelated position measurements the minimum detectable
change in velocity is $\delta v_{\rm min} \sim \delta \theta R_0/
\sqrt{N_{\rm obs}} T$, where $T$ is the time baseline over which the
orbit is monitored, and $N_{\rm obs}$ is the number of position
measurements taken in time $T$. Assuming $\delta \theta =
0.5 \trm{ mas}$ (\S~\ref{sec:astrometry}), $T = 10 \trm{ yr}$, and
$N_{\rm obs}=100$ yields $\delta v_{\rm min} \sim 0.2 \trm{ km
s}^{-1}$.  By equation (\ref{eq:deltav}) we have
\bea
\label{eq:bmax}
b_{\rm max} & = & b_0 \sqrt{ \left(\frac{2 m_i}{m_i + m_j}
\frac{v_{\rm rel}}{\delta v_{\rm min}} \right)^2 - 1 } \approx \frac{2
G m_i}{v_{\rm rel} \delta v_{\rm min}} , 
\eea 
where the approximation assumes $v_{\rm rel} \gg \delta v_{\rm min}$
and $m_j \lesssim m_i$. Assume $v_{\rm rel} \sim v_p =
\left(G M (1+e) / (1-e) a \right)^{1/2}$ for an encounter near
pericenter and $v_{\rm rel} \sim \left(G M / a \right)^{1/2}$ for an
encounter near apocenter. If we take $\delta v_{\rm min} \sim 0.2
\trm{ km s}^{-1}$ and $m_i = 10 M_\odot$ then for an S0-2 like orbit,
$b_{\rm max} \sim 10 \textrm{ AU}$ at pericenter and $b_{\rm max} \sim
50 \textrm{ AU}$ at apocenter.

We ignore the effect of the black hole on the encounter and treat the
interaction between the stars as a two-body problem. This is a fair
approximation as long as the duration of the encounter is much shorter
than the time scale over which the orbital velocity changes
significantly due to the influence of the black hole.  At pericenter
passage, where the orbital acceleration is greatest, the orbital time
scale is $t_p\sim (1-e)^{3/2} P$, where $P$ is the orbital period.
The two-body approximation is valid as long as the duration of the
encounter satisfies $t_{\rm enc} \sim b_{\rm max}/ v_{\rm rel} \ll
t_p$.  For an S0-2 like orbit $t_p \sim 0.5 \trm{ yr}$ while by
equation (\ref{eq:bmax}) $t_{\rm enc} \la 0.01 \trm{ yr}$ even for
$m_i = 20 M_\odot$.

Next, we estimate the rate at which encounters $b<b_{\rm max}$ occur
for a star on a given orbit (see, e.g., \citealt{Yu:03}).  Let
$\Gamma_{ij} (\vec{r},\vec{v}_j, t) \, dm_i$ be the rate at which a
star with mass $m_j$ at position $\vec{r}$ with velocity $\vec{v}_j$
at time t encounters stars with masses in the range range $m_i
\rightarrow m_i + dm_i$. Assume the number density of stars is
spherically symmetric and follows a power law $\nu(r) = \nu_0 (r /
r_h)^{-\alpha}$. The phase-space distribution function of the stars is
given by \citep{Magorrian:99} $f(\mathcal{E}) = h(\alpha)
\mathcal{E}^{\alpha - 3/2}$, where $\mathcal{E} = \Psi(r) - v^2/2$ and
$\Psi(r)$ is the relative gravitational potential at $r$, while
\beq
h(\alpha) = (2 \pi \sigma_h^2)^{-3/2} \nu_0 \frac{\Gamma(\alpha + 1)}{\Gamma(\alpha - 1/2)} 
             \sigma_h^{-2\alpha + 3},
\eeq
with $\sigma_h$ the linear stellar velocity dispersion outside
the sphere of influence of the BH $r_h \simeq 1 \trm{ pc}$.
The rate of detectable encounters in the mass bin is then
\bea
\label{eq:Gamma_j}
\Gamma_{ij}  & = & \int_0^\infty dv_i \frac{2 \pi v_i}{v_j} f(\vec{r}, v_i)
                      \int_{|v_i - v_j|}^{v_i + v_j} \, dv_{\rm rel} \, v_{\rm rel}^2 \,
		      \Sigma(v_{\rm rel}) \nonumber \\
            & = &  2 \pi K h(\alpha) \nonumber \\ & & \times 
           \int_0^{\sqrt{2 \Psi}}\left(\Psi - \frac{v_i^2}{2} \right)^{\alpha - 3/2}   \frac{v_i}{v_j} 
                 \left( v_i+v_j - |v_i-v_j| \right) dv_i, \nonumber \\ & &
\eea
where the cross section for detectable encounters $\Sigma = \pi b_{\rm
max}^2$, and $K = 4 \pi G^2 m_i^2 / \delta v_{\rm min}^2$.

We now determine the rate at which stars that will be monitored with
an ELT undergo detectable encounters. The integral in equation
($\ref{eq:Gamma_j}$) is most easily evaluated in the special case
$\alpha=3/2$, which is compatible with current observational
constraints \citep{Genzel:03a}.  To obtain a rough estimate of the
rates, consider the case $\alpha = 3/2$ and assume the background
stars all have identical mass not smaller than that of the test star
(e.g., they are a population of stellar mass black holes). By equation
(\ref{eq:Gamma_j})
\beq
\Gamma_j(\vec{r},\vec{v}_j,t) = \frac{3 \sqrt{2}}{8} \frac{K \nu_0}{\sigma_h^3} (2 \Psi - \frac{1}{3}v_j^2),
\eeq
and upon averaging over the orbital phase 
\bea
\Gamma_j(e,a) & = & \frac{1}{P}\int_0^P  \Gamma_j(\vec{r},\vec{v}_j, t) \,dt \nonumber \\
 & = & \frac{5 \pi \sqrt{2} \nu_0}{2\sigma_h^3} 
     \left(\frac{G m_i}{\delta v_{\rm min}}\right)^2 \frac{G M_{\rm bh}}{a}.
\eea
Assume the $N$ stars monitored with an ELT have an eccentricity
distribution uniform in $e^2$ (isotropic velocity ellipsoid) so that
$dN / de \, da \propto e a^{2-\alpha}=e\sqrt{a}$.  Integrating over
these distributions and normalizing to $N = N(<a_2)\propto
a_2^{3/2}$ yields the total rate at which encounters are detected with
an ELT
\bea
\label{eq:rate}
\Gamma(\alpha=3/2) & = & \int_0^1 \! \int_{a_1}^{a_2} \frac{dN(e,a)}{de \, da} 
\Gamma_j(e,a) \, de \, da \nonumber \\
     & = & \frac{15 \pi \sqrt{2} \nu_0}{2 \sigma_h^3} 
     \left(\frac{G m_i}{\delta v_{\rm min}}\right)^2 
  \frac{G M_{\rm bh} N}{a_1 + a_2 + \sqrt{a_1 a_2}},
\eea
where $a_1$ and $a_2$ define the range in semi-major axis that is
accessible to observations.

Given the above expression for the encounter rate for $\alpha=3/2$, we
rely on scaling relations to estimate the rate for different $\alpha$.
Since the encounter rate is proportional to the stellar density,
$\Gamma(\alpha) \simeq \Gamma(3/2) (r_h / r)^{\alpha - 3/2}$. Thus, if
$\alpha=7/4$, the rate of encounters is $\sim 3$ times larger than for
$\alpha = 3/2$.  The time scale for detectable encounters is therefore
\bea
\Gamma^{-1} & \sim & 0.3  \trm{ yr}  
 \left(\frac{m_i \nu_0}{2 \times 10^5 M_\odot\trm{ pc}^{-3}}\right)^{-1} 
 \left(\frac{m_i}{10 M_\odot}\right)^{-1}
\nonumber \\ & & \times \left(\frac{N}{100}\right)^{-1} 
 \left(\frac{a_2}{3000 \trm{ AU}}\right)
  \left(\frac{\sigma_h}{100 \trm{ km s}^{-1}}\right)^3 \nonumber \\ & & \times
 \left(\frac{\delta \theta}{0.5 \trm{ mas}}\right)^{2}  
 \left(\frac{T}{10 \trm{ yr}}\right)^{-2}
 \left(\frac{N_{\rm obs}}{100}\right)
 \left(\frac{a_2}{r_h}\right)^{\alpha - 3/2},\nonumber \\ & &
\eea
where we use the results of \S~\ref{sec:observations} that $N
\approx 100$, $a_1 \simeq 200 \trm{ AU}$, and $a_2 \simeq 3000 \trm{
AU}$, and have also assumed that the mass density of background
particles $m_i\nu_0$ is constant and independent of $m_i$.

Therefore, assuming a density cusp dominated by $\sim10M_\odot$ black
holes, $\sim 30$ nearby stellar encounters will be detectable during
ten years of monitoring with an ELT with $\delta \theta = 0.5 \trm{
mas}$.  Measurement of the frequency of detectable orbital deflections
$\Gamma\propto m_i$ is a direct test of the average mass of the dark
remnants that probably dominate the mass density near the black hole
\citep{Morris:93,Miralda:00} but are otherwise not directly
detectable. Since $N(<a) \propto a^{3/2}$, then by equation
(\ref{eq:rate}), $\Gamma \propto a^{1/2}$, i.e., the encounter rate
increases with distance from the massive black hole. The stars at $a >
3000 \trm{ AU}$ with detectable linear proper motion may therefore
yield the strongest constraint on the mass function of stellar
remnants, despite being below the threshold for detecting accelerated
motion due to the massive black hole.

\section{Method}
\label{sec:method}
In this section we describe how we generate mock ELT orbital data.
We also describe our implementation of the Markov Chain Monte Carlo
(MCMC) method which we use to estimate the uncertainties in the
orbital parameters and map the shape of the likelihood surface.

\subsection{Parameter Estimation}
\label{sec:MCMC}
We are interested in estimating the uncertainties in the parameters
given proper motion and radial velocity information for a sample of
$N$ stars orbiting the massive black hole at the GC. Each star's
projected orbit is described by six phase space parameters. The black
hole mass, its 3-dimensional position, and the normalization and slope
of the extended matter distribution, contribute an additional six
parameters. The entire parameter space of our model therefore has
dimension $J = 6N + 6$.

Parameter estimation on a $J$-dimensional grid is not practical. Since
the computational cost of the grid-based approach increases
exponentially with $J$, the parameter space becomes prohibitively
large for even just two or three stars. By contrast, the cost of the
MCMC method scales almost linearly with $J$.

We now briefly describe the basic ideas of the MCMC method and our
choice of implementation.  A general discussion of the theory and
application of the MCMC approach is given in \citet{Gilks:96}. Readers
not interested in the details of our parameter estimation scheme can
skip ahead to \S~\ref{sec:mockdata}.

Let $D$ denote the observed data, $\theta$ the model parameters,
$P(\theta)$ the prior distribution (which is uniform here), and $L(D |
\theta)$ the likelihood of detecting the data for a given set of
parameter values. By Bayes's theorem the distribution of $\theta$
conditioned on $D$ is given by
\beq
\label{eq:posterior}
\pi(\theta | D) = \frac{P(\theta) L(D | \theta)}{\int P(\theta) L(D | \theta) d\theta},
\eeq
and is called the posterior distribution of $\theta$. The statistical
properties of the parameters such as means, moments and confidence
contour levels, are entirely specified by $\pi(\theta | D)$.

Explicit evaluation of the integral in the denominator of equation
(\ref{eq:posterior}) is not practical in large dimensional models.
The MCMC method avoids evaluating the integral by instead generating a
Markov chain of parameter points the distribution of which converges
to the posterior distribution $\pi(\theta | D)$. The Markov aspect
refers to the property that the probability distribution of the $n$th
state (i.e., point) in the chain $\theta_n$ depends only on the
previous state $\theta_{n-1}$. It can be shown (e.g.,
\citealt{Gilks:96}) that the density of points in a Markov chain
converges to $\pi(\theta | D)$ if the following criteria are
satisfied: (1) the chain is irreducible, namely from any starting
state $\theta_0$ the chain can reach any non-empty set with positive
probability in some finite number of iterations; (2) the chain is
aperiodic in that it does not oscillate between different sets of
states in a regular periodic fashion; (3) the chain is positive
recurrent, meaning that if the initial value $\theta_0$ is sampled
from the posterior then the expected time (i.e., number of iterations)
to return arbitrarily close to state $\theta_0$ is finite. There are
several algorithms for generating Markov chains that satisfy the above
properties. We use the Metropolis algorithm \citep{Metropolis:53} in
our numerical calculations.

Our implementation of the Metropolis algorithm is as follows.
\begin{enumerate}
\item Start a chain at $t=0$ with some initial state $\theta_0$.
\item Generate a trial state $\theta'$ according to the jump proposal
      distribution $q(\theta' | \theta_t)$ (see below). Compute
\beq
\alpha(\theta_t, \theta')  =  \min \left[1, \frac{L(D | \theta')}{L(D | \theta_t)}\right] .
\eeq
\item Sample a uniform random variable $U$ that lies between $(0,1)$.
\item If $U \le \alpha(\theta_t, \theta')$ then set $\theta_{t + 1} =
  \theta'$ (i.e., accept the jump). If $U > \alpha(\theta_t, \theta')$
  then set $\theta_{t + 1} = \theta_t$.
\item Increment $t$.
\item Go to step \#2.
\end{enumerate}
If the observational errors follow a normal
distribution, $L(D | \theta) \simeq \exp[-\chi^2(\theta, D)/2]$. 
The $\chi^2(\theta, D)$ statistic for a single star is given by
\bea
\label{eq:chisq}
\chi^2(\theta, D) & = & \sum_{i=1}^{M} 
\left\{\frac{[x_i(\theta) - x_i(D)]^2}{\sigma^2_{x, i}} + 
      \frac{[y_i(\theta) - y_i(D)]^2}{\sigma^2_{y, i}} \right\}\nonumber\\
          & & +   
\sum_{j=1}^{K} \frac{[v_j(\theta) - v_j(D)]^2}{\sigma^2_{v, j}},
\eea
where $(x, y)$ is the astrometric position of the star, $v$ its radial
velocity, and $\sigma$ the corresponding measurement errors (i.e.,
$\sigma_{x, y} = \delta \theta$ and $\sigma_v = \delta v$).  We
simultaneously fit to multiple stars by summing each star's $\chi^2$
to form a cumulative $\chi^2$ for the model.

The jump proposal distribution $q(\theta' | \theta_t)$ is the
probability of selecting a trial state $\theta'$ given the current
state $\theta_t$. For the Metropolis algorithm one considers only
symmetric proposals of the form $q(\theta' | \theta_t) = q(\theta_t |
\theta')$. We choose to model the jump distribution as a multivariate
normal distribution with mean $\theta_t$ and constant covariance
matrix $C$.

Although the distribution of points in a chain is independent of the
form of the jump distribution once the Markov chain has converged, the
time it takes a chain to converge is sensitive to the jump
distribution. To ensure an efficient run one must carefully chose the
shape and step size of the jump distribution. An ideal jump
distribution has a shape and step size that not only minimizes the
convergence time but also samples the entire posterior distribution
efficiently. In our implementation the shape of the jump distribution
is determined by $C$ and the step size is determined by a constant
scale factor multiplying $C$.

$C$ is chosen such that the shape of the jump distribution is similar
to that of the posterior distribution, although we again emphasize
that the shape is only important for the efficiency of
convergence. This ensures that the chain mixes well even in regions of
degeneracy. To this aim, we compute the covariance matrix that
describes the shape of the $\chi^2$ surface in the neighborhood of its
minimum.  We first compute the approximate best fit parameter state
$\theta_{\rm bf}$ by minimizing $\chi^2$.  We then specify a pilot
covariance matrix $C_p$ that is purely diagonal with variances given
by a reasonable guess of the $1 \sigma$ uncertainties for individual
parameters. We draw a number ($\sim 1000$) of pilot points from a
multivariate normal distribution with mean $\theta_{\rm bf}$ and
covariance $C_p$. Since the pilot points are within $\sim 1 \sigma$ of
the $\chi^2$ minimum, the shape of the $\chi^2$ surface in the region
of the points is approximately quadratic. We solve the linear
least-squares problem by fitting a quadratic $\chi^2$ model to the
points and obtain the approximate Fisher matrix that describes the
curvature of the $\chi^2$ surface. We then determine the eigenvalues
and eigenvectors of the Fisher matrix.  If any of the eigenvalues are
negative, indicating that the shape of the surface is unconstrained in
some direction, we generate more pilot points and redo the linear
least-squares fit. Finally, we invert the resulting Fisher matrix to
obtain the covariance matrix $C$.

The constant scale factor that determines the step sizes must also be
carefully chosen to ensure that the chain effectively explores the
parameter space. If the steps are too small the chain does not mix
well as it stays in one region of the parameter space for long periods
of time. If the steps are too large, the trial states are rejected
frequently. For a multivariate normal jump distribution the most
efficient step sizes are those for which $\sim25\%$ of the jump
proposals are accepted (see \citealt{Gelman:95}).  We chose the
(constant) jump scale factor to optimize the acceptance rate.

The first steps in a chain may be sensitive to the starting state
$\theta_0$ and are therefore not sampled from the posterior
distribution. We discard these initial ``burn in'' points. To ensure
that a chain has converged and is sampling the full posterior
distribution we run multiple chains each starting at widely dispersed
states. We tested for convergence with the Gelman-Rubin test statistic
\citep{Gilks:96}.

\subsection{Mock Data}
\label{sec:mockdata}
\begin{deluxetable}{llllll}
\tablewidth{0pt}
\tablecaption{Stellar Orbital Parameters}
\tablehead{
	   \colhead{Star}&
	   \colhead{$P$} &
	   \colhead{$a$} &         
           \colhead{$e$}  &
	   \colhead{$r_{\rm min}$} &
	   \colhead{$i$} \\
	   \colhead{}&
	   \colhead{(yr)} &
	   \colhead{(AU)} &
           \colhead{}  &
	   \colhead{(AU)} &
	   \colhead{(deg)} 	  
}
\startdata
\label{tab:stars}
 1 &  2.9 &  325 & 0.852 &  48 &  89  \\
 2 &  5.2 &  476 & 0.587 & 197 & 104 \\
 3 &  5.4 &  488 & 0.783 & 106 &  46  \\
 4 &  9.5 &  712 & 0.560 & 313 &  86  \\
 5 & 13.4 &  895 & 0.800 & 179 & 114 \\
 6 & 14.2 &  931 & 0.772 & 212 &  81  \\
 7 & 17.7 & 1078 & 0.800 & 216 &  15  \\
 8 & 23.2 & 1289 & 0.605 & 510 & 126 \\
 9 & 25.2 & 1363 & 0.903 & 132 &  25  \\
10 & 26.6 & 1413 & 0.730 & 382 & 142 \\
11 & 36.8 & 1755 & 0.705 & 517 &  56  \\
12 & 40.4 & 1869 & 0.890 & 206 &  93  \\
13 & 42.3 & 1928 & 0.885 & 223 & 105 \\
14 & 46.6 & 2056 & 0.703 & 611 &  42  \\
15 & 50.6 & 2172 & 0.552 & 973 & 104 \\
16 & 51.5 & 2197 & 0.567 & 951 & 145 \\
17 & 52.2 & 2218 & 0.439 &1245 & 106 \\
18 & 52.5 & 2225 & 0.636 & 810 &  95  \\
19 & 57.8 & 2372 & 0.319 &1617 &  26  \\
20 & 78.1 & 2901 & 0.220 &2264 &  60  \\
\enddata
\tablecomments{ The listed parameters are: orbital period ($P$), 
semi-major axis ($a$), eccentricity ($e$), pericenter distance ($r_{\rm min}$),
and inclination ($i$).  They are not all independent variables.}
\end{deluxetable}

To generate a realistic set of orbital data we must determine: (i) the
number of stars $N$ we can detect and monitor with an ELT, (ii) the
spatial distribution of these stars, (iii) the number of observations
per year per star and, (iv) the observational errors in the stellar
positions and velocities. In \S~\ref{sec:observations} we showed that
with a 30 meter ELT the position of the stars can be centroided to an
astrometric precision $\delta \theta_{30}$ between $0.1 - 0.5 \trm{
mas}$ and the radial velocities measured to accuracies $\delta v_{30}$
between $1 - 10 \trm{ km s}^{-1}$.  We found that with such an ELT we
can detect the accelerated proper motion of approximately 100
stars. We estimate that an integral-field spectrograph on an ELT
enables a dedicated GC observing program to obtain position and
velocities of each of the 100 stars roughly ten times per year.

A realistic mock data set might therefore consist of $N = 100$ stars,
observed over a ten year baseline with ten observations per year per
star, with position and velocity measurements for each star accurate
to $0.5 \trm{ mas}$ and $10 \trm{ km s}^{-1}$. Unfortunately, running
our MCMC simulation on such a large data set was not feasible due to
limits in computational speed. A typical run requires $\sim 10^7$
iterations (i.e., jumps) in order to fully sample the posterior
distribution. This corresponds to a minimum of $\sim 3$ days on a
desktop machine for just 20 stars ($J \simeq 126$) with 100 points per
star; a simulation with 100 stars takes approximately five times
longer. However, one can obtain realistic results from a reduced
sample size by properly scaling the $\chi^2$ values (see equation
\ref{eq:chisq}) to emulate the full sample size. In particular, we
construct a mock data set with $N=20$ and multiply the $\chi^2$ of
each star by a factor of five.

\begin{figure}
\begin{center}
\epsscale{1.2}
\includegraphics[bb= -30 -20 580 501, angle=0,scale=0.45]{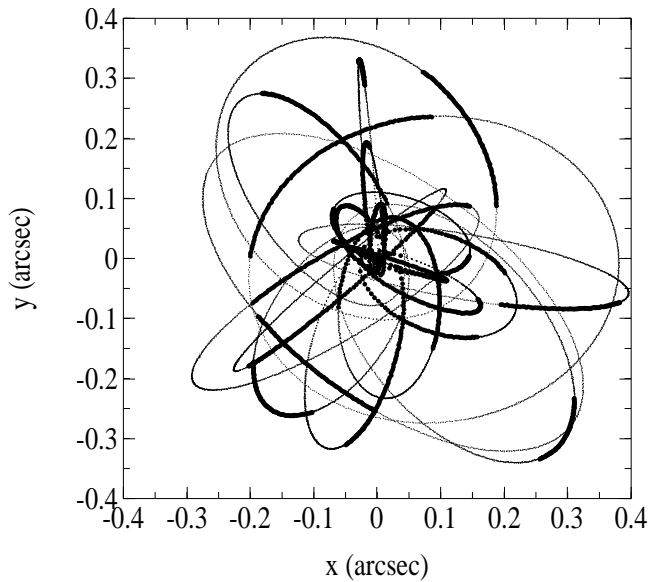}
\caption{Astrometric positions of the 20 synthesized orbits to which
we fit the model. The thick-lined portion of each orbit is the proper
motion over the fitted 10 year baseline assuming 10 observations per
year.\label{fig:orbits}}
\end{center}
\end{figure}

This approach yields realistic estimates of parameter uncertainties as
long as the mock data set with $N=20$ stars fairly represents the full
data set with 100 stars. We minimize the effects of sample variance as
follows. We first generate data for 1000 synthetic orbits. These
orbits are drawn from the distribution function of the power-law
density profile assuming randomly oriented orbits and considering only
those orbits with semi-major axes in the range detectable with an ELT
(see \S~\ref{sec:observations}). We generate mock data for these
orbits assuming Gaussian position and velocity errors with dispersions
$\delta \theta$ and $\delta v$ and a specific input model for the
potential (e.g., black hole plus extended matter).  For each
individual star, we compute the difference in $\chi^2$ between the
best-fit model (essentially the model used to generate the data) and
the null hypothesis model (e.g., no extended matter). We then rank the
stars by the size of this $\chi^2$ difference. We bin the 1000 stars
into $N$ bins according to their rank and randomly select one star
from each bin. The resulting $N$ stars form the set of orbits to which
we fit.

Table \ref{tab:stars} lists the orbital parameters for one realization
of a sample of 20 stars to which we fit. Given the orbital parameters,
we generate mock data by solving the equation of motion for each star
(see equation \ref{eq:EOM}). In Figure \ref{fig:orbits} we show the
astrometric positions of the 20 stars over the ten year observational
baseline with ten epochs per year. The values of the input model
parameters describing the potential are: $M_{\rm bh} = 4 \times 10^6
M_\odot$, $R_0 = 8 \trm{ kpc}$, $(x_{\rm bh}, y_{\rm bh}) = (0, 0)$,
$M_{\rm ext}(r < 0.01 \trm{ pc}) = 6000 M_\odot$ and either $\gamma =
1.5$ or 2.

To test that the parameter uncertainty estimates are not affected by
sample variance we ran simulations on several different draws of 20
stars. As we show in \S~\ref{sec:results} the parameter uncertainties
obtained are similar amongst the different data sets, suggesting that
sample variance does not affect the results. Thus, given the current
uncertainties in an ELT's ultimate capabilities as well as the
uncertainty in the exact nature of the stellar distribution at the GC,
we conclude that to a reasonable approximation a mock data set
comprised of $N = 20$ stars with $\chi^2$ values increased five-fold
yields parameter uncertainties similar to that expected with
observations by a 30 meter ELT.

We also show in the next section that the orbital parameter
constraints scale with the measurement errors $\sigma$ and number of
stars $N$ as $\sigma / N^{1/2}$. Results for a wide range of assumed
ELT capabilities (i.e., different $\delta \theta$ and $\delta v$,
different aperture, etc.) can therefore be computed by scaling the
results of our fiducial 30 meter ELT model, using the relations
between $N$, aperture $D$, $\delta \theta$, and $\delta v$, given in
\S~\ref{sec:observations}.

\section{Results}
\label{sec:results}

\begin{figure}[t!]
\epsscale{1.0}
\includegraphics[bb= -40 -20 580 501, angle=0,scale=0.45]{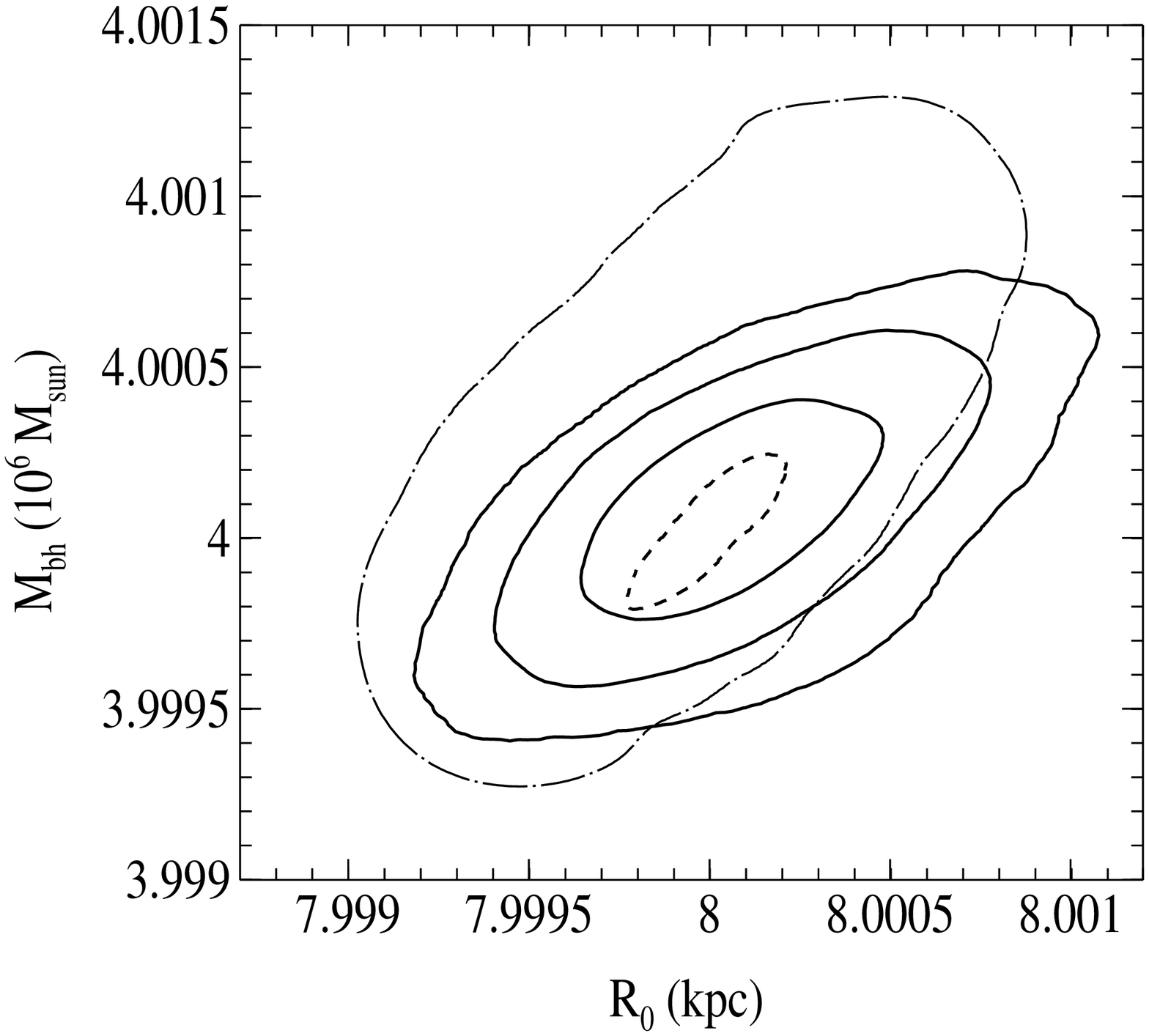}
\caption{The constraints on $M_{\rm bh}$ and $R_0$ obtainable with a
an ELT assuming an extended matter distribution with $\gamma = 1.5$
(results are similar for $\gamma = 2$). The solid contours show the
68\%, 95\%, and 99.7\% confidence levels assuming an astrometric limit
of $\delta \theta = 0.5 \trm{ mas}$ and a spectroscopic limit of
$\delta v = 10 \trm{ km s}^{-1}$ for the draw of 20 stars shown in
Table \ref{tab:stars}. The line-dot contour shows the 99.7\%
confidence level for a different draw of 20 stars. The dashed contour
shows the 99.7\% confidence level for smaller astrometric and
spectroscopic limits of $\delta \theta = 0.1 \trm{ mas}$ and
$\delta v = 2 \trm{ km s}^{-1}$.  An ELT will constrain both $M_{\rm
bh}$ and $R_0$ to better than $0.1\%$.
\label{fig:ngltmr}}
\end{figure}
\begin{figure}[t!]
\epsscale{1.0}
\includegraphics[bb= -40 -20 580 501, angle=0,scale=0.45]{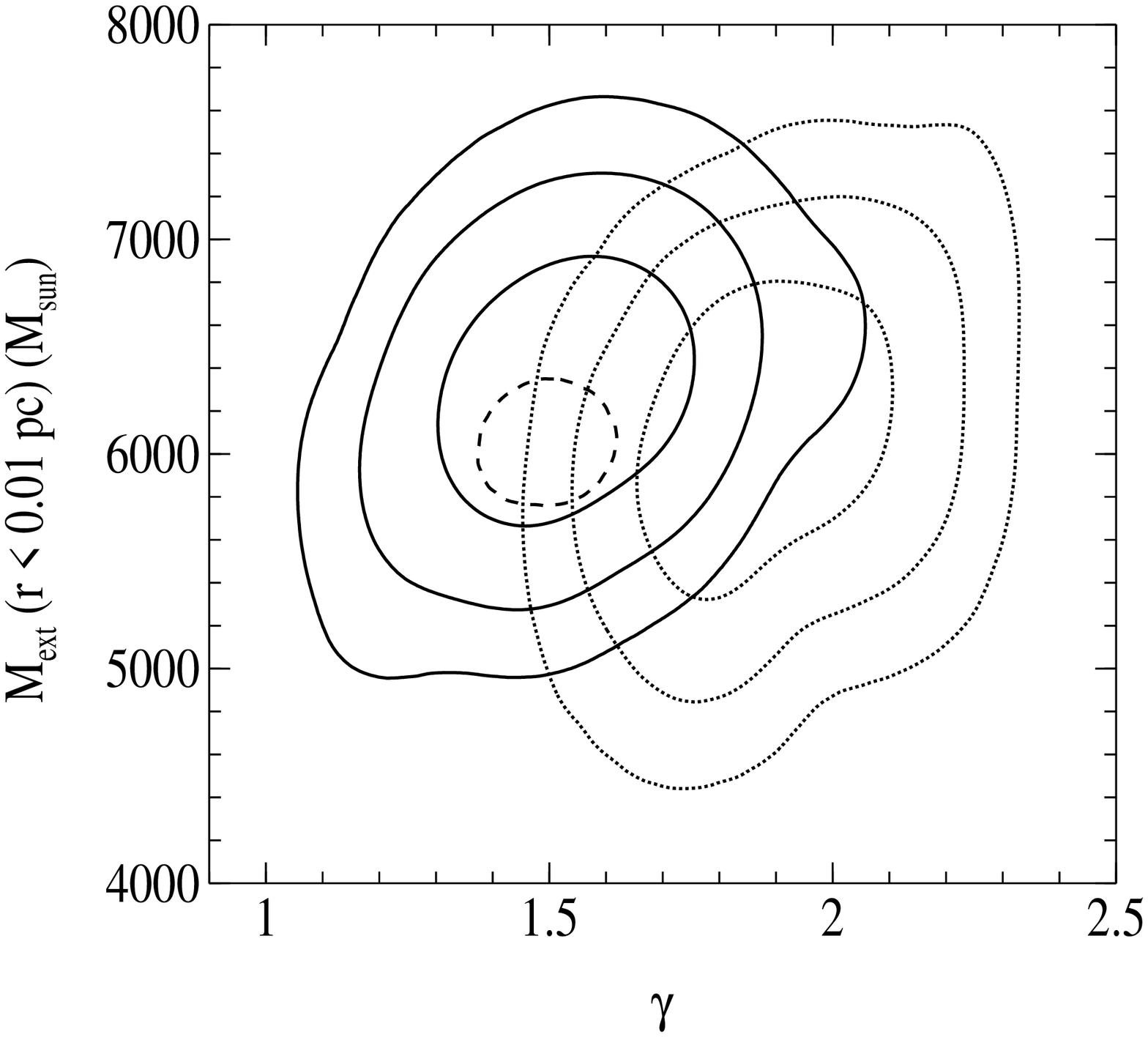}
\caption{The constraint on the extended matter distribution obtainable
with an ELT. Shown are the $68\%$, $95\%$, and $99.7\%$ confidence
levels on the enclosed mass and slope of an extended matter
distribution assuming an astrometric limit of $\delta \theta = 0.5
\trm{ mas}$ and a spectroscopic limit of $\delta v = 10 \trm{ km
s}^{-1}$. The input models have power-law slope of $\gamma = 1.5$ and
$\gamma = 2$ and an input enclosed mass of $6000 M_\odot$ within $0.01
\trm{ pc}$. The dashed contour is the constraint at the 99.7\% level
for measurement errors that are a factor of five
smaller.\label{fig:ngltextended}}
\end{figure}

In this section we investigate how well observations with an ELT
constrain the structure of the GC.  Our model of the GC and the orbits
was described in \S~\ref{sec:orbitmodel}.  We draw stellar orbital
parameters from a phase-space distribution determined by the model and
use these orbits to synthesize mock ELT data (see
\S~\ref{sec:method}). We then fit a model to the mock data and
calculate the uncertainties in the parameters using the MCMC technique
discussed in \S~\ref{sec:MCMC}. We show results for a 30 meter
ELT with $(\delta \theta, \delta v) = (0.5 \trm{ mas}, 10 \trm{ km
s}^{-1})$ and $(\delta \theta, \delta v) = (0.1 \trm{ mas}, 2 \trm{ km
s}^{-1})$. However, since the parameter uncertainties scale with
measurement error $\sigma$ and number of monitored stars $N$ as
$\sigma / N^{1/2}$, the results can be used to describe the
capabilities of an ELT with different specifications. For example, a
100 meter ELT will detect $\sim 10 \times$ as many stars
(\S~\ref{sec:confusion}); if the telescope has astrometric and
spectroscopic errors that are smaller than those of a 30 meter
telescope by a factor of five the parameter uncertainties will be
$\sim 10 \times$ smaller. In this section, we estimate the limits that
can be placed on the parameters associated with the black hole
including $M_{\rm bh}$ and $R_0$ (\S~\ref{sec:resultmr}), as well as
on the extended distribution of (dark) matter near the black hole
(\S~\ref{sec:resultextended}).  We discuss the dependence of the
limits on the astrometric and spectroscopic precision of the
observations.  We also investigate whether relativistic corrections to
the Keplerian motion can be detected at the GC
(\S~\ref{sec:resultsrelativ}).

\subsection{Measuring $M_{\rm bh}$ and $R_0$}
\label{sec:resultmr}
\begin{figure}
\begin{center}
\epsscale{1.0}
\includegraphics[bb= -40 -20 580 501, angle=0,scale=0.45]{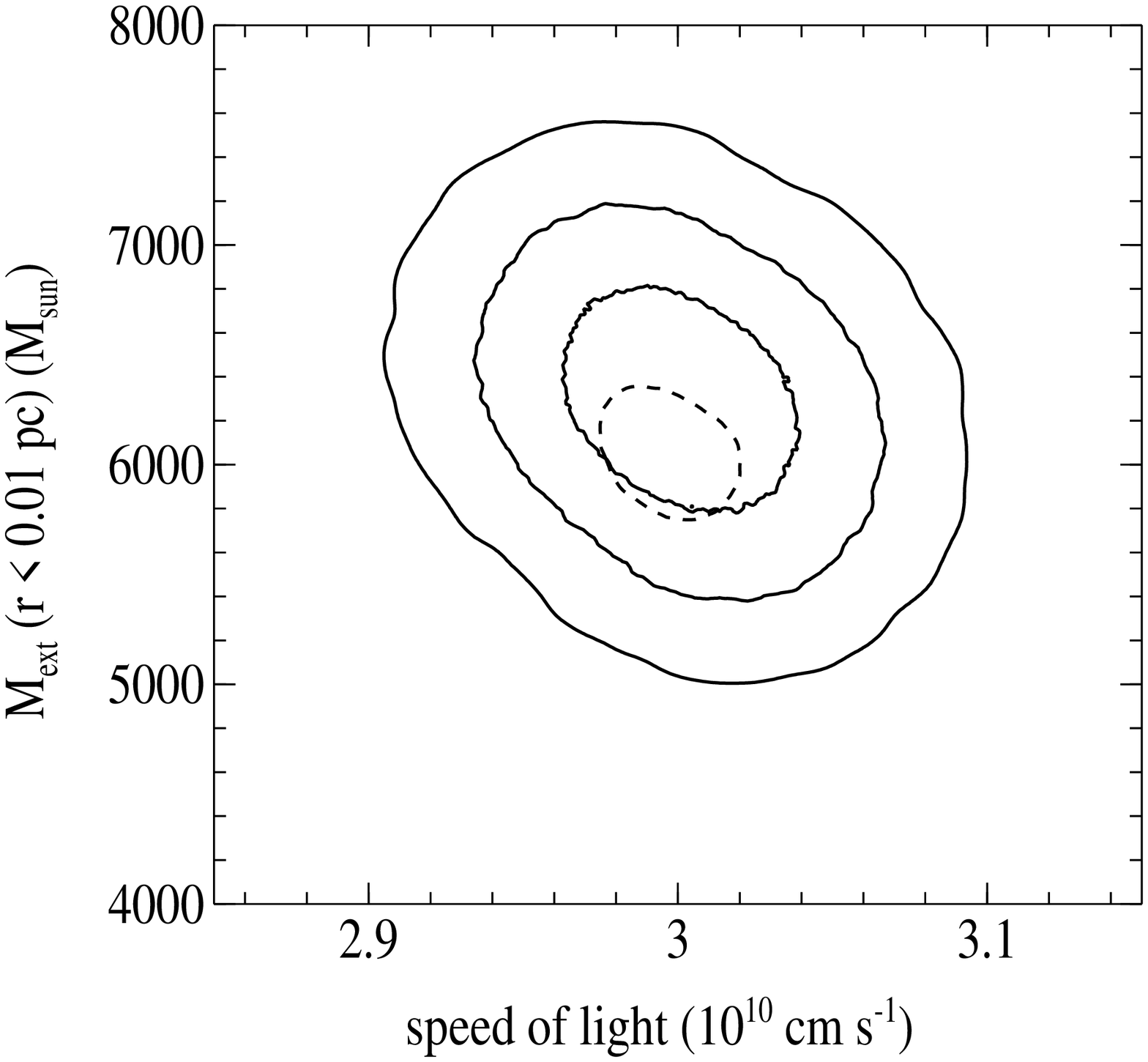}
\caption{An ELT's sensitivity to post-Newtonian effects assuming an
astrometric limit of $\delta \theta = 0.5 \trm{ mas}$ and a
spectroscopic limit of $\delta v = 10 \trm{ km s}^{-1}$. Shown is the
uncertainty in the speed of light and the extended matter mass as
obtained by including post-Newtonian corrections to the equations of
motion. The Roemer delay and special relativistic effects are not
included in the model in order to demonstrate that general
relativistic effects of order $(v/c)^2$, including the prograde
precession, are detectable with an ELT. The line styles are the same as
in Figure \ref{fig:ngltmr}.\label{fig:relativistic}}
\end{center}
\end{figure}
In Figure \ref{fig:ngltmr} we show the constraints an ELT will place
on $M_{\rm bh}$ and $R_0$. For an astrometric limit of $\delta \theta =
0.5 \trm{ mas}$ and a spectroscopic limit of $\delta v = 10 \trm{ km
s}^{-1}$ (see \S~\ref{sec:astrometry}) the fractional uncertainties in
$M_{\rm bh}$ and $R_0$ are less than 0.1\% at the 99.7\% level. This
is a factor of $\sim 100$ times better than present uncertainties. The
result is robust in that simulations with distinct mock data sets of
20 stars, drawn in the fashion described in \S~\ref{sec:mockdata},
produce similar uncertainties in the model parameters.

For astrometric and spectroscopic limits that are a factor of five
smaller the fractional uncertainties in $M_{\rm bh}$ and $R_0$ are
smaller by almost a factor of five. The uncertainties in $M_{\rm bh}$
and $R_0$ scale almost linearly with the measurement errors for
observations at this precision. We also verified that the
uncertainties scale with $N$ as roughly $N^{-1/2}$.

Observations with a 30 meter ELT will therefore constrain the distance to the
GC to within a few parsecs and the mass of the black hole to within a
few thousand solar masses. We discuss the implications of measuring
$R_0$ to such high accuracy in \S~\ref{sec:discussion}.

\subsection{Measuring the Extended Matter Distribution}
\label{sec:resultextended}
In Figure \ref{fig:ngltextended} we show the constraints an ELT will
place on the extended matter distribution for input power-law models
with $M_{\rm ext} (r < 0.01 \trm{ pc}) = 6000 M_\odot$ and $\gamma =
1.5$ or $\gamma = 2$. We chose these distributions in order to conform
to the extrapolation of the observed stellar density distribution and
to theoretical estimates of dark matter clustering (see
\S~\ref{sec:intro}). We find that for an ELT with $\delta \theta = 0.5
\trm{ mas}$ and $\delta v = 10 \trm{ km s}^{-1}$ one can detect such
extended matter distributions, yielding measurements of $M_{\rm ext}$
and $\gamma$ that are accurate to $20 - 30\%$ (i.e., $\delta M_{\rm
ext} \sim 1500 M_\odot$ and $\delta \gamma \sim 0.5$).  Since the
amplitude of the Newtonian retrograde precession varies linearly with
$M_{\rm ext}$ (\S~\ref{sec:newtonian}), the fractional uncertainty is
$\delta M_{\rm ext} / M_{\rm ext} \propto \delta \Delta \phi_{\rm
Newt} / \Delta \phi_{\rm Newt} \propto M_{\rm ext}^{-1}$, where
$\delta \Delta \phi_{\rm Newt}$ is set by the astrometric
precision. Thus $\delta M_{\rm ext}$ is independent of $M_{\rm ext}$
so that an extended matter distribution is detectable (i.e.,
observations yield a lower bound) for $\delta \theta = 0.5 \trm{ mas}$
and $\delta v = 10 \trm{ km s}^{-1}$ as long as $M_{\rm ext} (r < 0.01
\trm{ pc}) \ga \delta M_{\rm ext} \sim 1500 M_\odot$.  Such an ELT
will therefore place interesting constraints on the extended matter at
the GC.

\subsection{Measuring Relativistic Effects}
\label{sec:resultsrelativ}
 
As discussed in \S~\ref{sec:relativistic}, order of magnitude
estimates suggest that post-Newtonian corrections to the equations of
motion, involving terms of order $(v/c)^2$, are measurable with an
ELT with astrometric resolution of $\delta \theta \la 0.5 \trm{
mas}$.  In an effort to demonstrate this more quantitatively, we allow
the speed of light to be a parameter in our model and examine how well
we recover its value. We purposely do not include relativistic
corrections to the observed motion associated with propagation effects
(e.g., the Roemer time delay and other higher-order corrections) so
that we can examine the detectability of $(v/c)^2$ general
relativistic corrections to the orbital dynamics such as the prograde
precession of the major axis position. In Figure
\ref{fig:relativistic} we show the constraint on $c$ as a function of
$M_{\rm enc}$. Post-Newtonian effects are observable, as $c$ is
measured to $\sim 5 \%$ accuracy. Since $v /c \la 0.2$ for all stars
in the sample (\S~\ref{sec:EOM}) the few percent constraint on $c$
suggests that while the $(v/c)^2$ effects are measurable, the
$(v/c)^3$ effects are not. The orbital precession due to black hole
spin is of order $(v/c)^3$ (\S~\ref{sec:spin}) and detecting it with
an ELT with $\delta \theta = 0.5 \trm{ mas}$ requires the favorable
discovery of a star on a compact and highly eccentric orbit. Based on
estimates of the signal-to-noise from a spin-induced apocenter shift
(equation [\ref{eq:SNspin}]), an astrometric precision of $\sim 0.05
\trm{ mas}$ is needed to reliably detect the black hole spin.

The degeneracy between $c$ and $M_{\rm enc}$ is a consequence of the
degeneracy between the prograde relativistic precession and the
retrograde Newtonian precession. Decreasing $c$ increases the amount
of prograde motion $\Delta \phi_{\rm pro}$ while increasing $M_{\rm
enc}$ increases the amount of retrograde motion $\Delta \phi_{\rm
Newt}$. The two effects compensate for one another over a range of $c$
and $M_{\rm enc}$. The degeneracy is broken at sufficiently extreme
values of $M_{\rm enc}$ because the relativistic and Newtonian effects
each induce a distinct precessional shape.

\section{Constraints on Galactic Structure from Measurements of $R_0$}
\label{sec:discussion}
The distance to the GC, $R_0$, is a fundamental parameter in models of
the Milky Way structure. As \citet{Olling:01} note, models of the
Milky Way exhibit strong interrelations between the Galactic constants
($R_0$ and the local Galactic rotation speed $\Theta_0$), the
shortest-to-longest axis ratio, $q = c / a$, of the dark matter halo,
and the local stellar column density $\Sigma_\ast$.  The determination
of $q$ is of particular interest since different models of dark-matter
and structure formation scenarios predict different values for
$q$. Cold dark-matter simulations typically produce galactic halos
that are triaxial \citep{Warren:92,Jing:02} although these tend 
become oblate under the influence of the dissipative infall of gas
resulting in halos with $q \simeq 0.5$
\citep{Dubinski:94}. Alternatively, hot dark-matter models predict
round halos with $q \sim 0.8$ \citep{Peebles:93} while some baryonic
dark matter models imply $q \sim 0.2$ \citep{Pfenniger:94}. As we now
discuss, determining $R_0$ to $0.1 \%$ via monitoring of stellar
orbits at the GC with an ELT enables an extremely precise measurement
of $q$ in the Milky Way.

\citet{Olling:00} demonstrate that there is significant uncertainty in
existing estimates of $q$ in galaxies due to both the limited amount
of data available for measuring $q$ and the fact that different
measurement techniques have yielded systematically different values.
Presently, the situation is not any better for our own Galaxy, with
plausible values lying in the range $0.3 \la q \la 1$.

The measurement of $q$ in the Milky Way entails measuring the Galaxy's
radial mass distribution and the degree to which this mass
distribution is flattened. \citet{Olling:00} show that the uncertainty
in $q$ in the Milky Way is almost entirely due to the large errors in
the Galactic constants $\Theta_0$ and $R_0$. Indeed, \citet{Olling:01}
show that the fractional uncertainty in $q$ is nearly twice the
fractional uncertainty in $\Theta_0$.  Therefore, a precision
measurement of the Sun's proper motion with respect to the GC in
combination with a precision measurement of $R_0$ tightly constrains 
$\Theta_0$ and hence $q$.  According to \citet{Salim:02}
future astrometric surveys will be able to measure the Sun's proper
motion $\mu = V / R_0$ to within several microarcseconds,
corresponding to $0.1 \%$ accuracy. Here $V = \Theta_0 + V_\odot
\simeq 220 \trm{ km s}^{-1}$ is the sum of the rotation speed of the
local standard of rest and the Sun's motion relative to it.  The
uncertainty in $\Theta_0$ will be the dominant error in $V$; $V_\odot$
is already known to an accuracy of $0.6 \trm{ km s}^{-1}$ from the
\emph{Hipparcos} catalogue \citep{Dehnen:98}. Thus, the monitoring of
stellar orbits at the GC with an ELT in conjunction with future
astrometric survey missions will constrain the Milky Way's dark matter
halo shape parameter $q$ to a few tenths of a percent.

\section{Conclusions}
\label{sec:conclusions}

We have examined a variety of experiments that can be achieved through
the infrared monitoring with an ELT of stars within a few thousand AU
of the GC.  The astrometric limit of a 30 meter ELT is conservatively
0.5 mas and possibly as high as 0.1 mas. By comparison, the
astrometric limit of current observations is $1 - 2 \trm{ mas}$.

The greater point-source sensitivity and spectral resolution of an ELT 
enables the measurement of radial velocities with errors $\la 10
\trm{ km s}^{-1}$.  At present, of the $\sim 10$ stars with measured
accelerated proper motions, spectral lines have been detected only in
S0-2, with radial velocity uncertainties of $\sim 30 \trm{ km
s}^{-1}$. Measuring the radial velocities of stars breaks the
degeneracy between mass and distance and thus yields a direct
measurement of the distance to the GC. If the spectra of fainter stars
can be obtained, the detection of deep molecular lines will improve
upon the velocity estimates by an additional factor $\times10$. The
solar type stars that will be detectable with an ELT may therefore
yield radial velocity uncertainties considerably smaller than $10
\trm{ km s}^{-1}$.

A 30 meter ELT will be able to detect stars down to a $K$-band
magnitude of $K \sim 22$, approximately four magnitudes fainter than
currently possible. Due to confusion, it will be difficult to detect
still fainter stars. Using measurements of the $K$-band luminosity
function within the inner $1\arcsec$ of the GC, we estimate that such
an ELT will detect the accelerated motion of $\sim 100$ stars with
semi-major axes in the range $200 \la a \la 3000 \trm{ AU}$. Current
observations are limited to the detection of $\sim 10$ stars, all with
$a \ga 1000 \trm{ AU}$. We find that the number of stars with
detectable accelerated motion scales with the aperture of an ELT as
$N \simeq 100 (D / 30 \trm{ m})^2$.

Given the observational capabilities of an ELT and the likely, albeit
at low masses largely uncertain, stellar environment at the GC, we
constructed a plausible sample of stellar orbits. The model includes
the dynamical contribution of an extended distribution of dark matter
around the black hole that is composed of stellar remnants and CDM. We
find that for measurements at the precision obtainable with an ELT
the uncertainty in the model parameters scale with the measurement
errors $\sigma$ (i.e., $\delta \theta$, $\delta v$) and the number of
monitored stars $N$ as roughly $\sigma / N^{1/2}$. Thus, while we
focus on the capabilities of a diffraction limited 30 meter ELT with
$\delta \theta = 0.5 \trm{ mas}$ and $\delta v = 10 \trm{ km s}^{-1}$,
our results can be used to determine the capabilities of an ELT with
different specifications. For example, a 100 meter ELT will detect
$\sim 10 \times$ as many stars so that if it has astrometric and
spectroscopic errors that are smaller by a factor of five, the
measurement accuracy in the parameters will improve by a factor of
approximately ten.

We find that with a 30 meter ELT the parameters $M_{\rm bh}$ and
$R_0$ will be measured to an accuracy better than $0.1\%$. Determining
$R_0$ to within a few parsecs will significantly constrain models of
the Galactic structure as it aids the precise measurement of the dark
matter halo shape.

While current observations of stellar proper motions are compatible
with Keplerian motion, a number of dynamical effects produce
significant deviations, including the Newtonian retrograde precession,
the relativistic prograde precession, frame dragging due to the black
hole spin, and interstellar interactions involving nearby encounters.
All but the frame dragging effect produce non-Keplerian motions that
are detectable with a 30 meter ELT.  Unfortunately, the spin of the massive
black hole at the GC will probably be out of reach to kinematic
studies unless an astrometric precision of $\sim 0.05
\trm{ mas}$ is achieved.

The presence of an extended distribution of matter results in a
Newtonian retrograde precession due to differences in the amount of
mass enclosed within an orbit's pericenter and apocenter. We
considered extended matter density profiles consistent with current
observations of the stellar distribution at the GC. We modeled the
distribution as a power-law profile normalized such that $M_{\rm
ext}(r < 0.01 \trm{ pc}) = 6000 M_\odot$ and with slope $\gamma = 1.5$
or 2. Standard models of dark matter clustering about a massive black
hole predict similar profiles. An orbit monitoring program with a 30 meter ELT 
will constrain the mass and slope of such profiles to $\sim 30\%$
accuracy. Thus, monitoring orbits with an ELT provides a probe of the
extended matter distribution within $\sim 10^4$ Schwarzschild radii of
the massive black hole at the GC.

We also calculated the rate at which the monitored stars experience
detectable deflections due to stellar gravitational scattering
encounters with background compact remnants. We considered a detection
threshold set by the minimum detectable change in the velocity of a
monitored star.  For a density cusp dominated by $\sim 10 M_\odot$
black holes, $\sim 30$ nearby stellar encounters will be detected by 
a 30 meter ELT over a ten year observing baseline.  This will confirm the 
presence of a cusp of compact remnants at the GC and enable the measurement of the
remnants' masses.

\acknowledgements 

The authors would like to thank M.~Kamionkowski, D.~Figer,
K.~Matthews, and E.~Pfahl for comments and E.~Agol, S.~Phinney, and
J.~Graham for helpful discussions. We also thank the referees for
their helpful comments. NNW acknowledges the support of an NSF 
graduate fellowship and DoE DE-FG03-92ER40701.  MM was supported 
at Caltech by a postdoctoral fellowship from the Sherman Fairchild Foundation.

\end{document}